\documentclass[conference,compsoc]{IEEEtran}
\ifCLASSOPTIONcompsoc
  \usepackage[nocompress]{cite}
\else
  \usepackage{cite}
\fi
\ifCLASSINFOpdf
\else
\fi

\usepackage{caption}
\usepackage{color}
\usepackage{mathtools}
\usepackage{booktabs}
\usepackage{algorithm}
\usepackage{dblfloatfix}
\usepackage{algpseudocode}
\floatname{algorithm}{Procedure}

\begin{document}
%
\title{Metric Factorization: Recommendation beyond Matrix Factorization}



%
\author{\IEEEauthorblockN{Shuai Zhang\IEEEauthorrefmark{1},
Lina Yao\IEEEauthorrefmark{1},
Yi Tay\IEEEauthorrefmark{2},
Xiwei Xu\IEEEauthorrefmark{3},
Xiang Zhang\IEEEauthorrefmark{1} and
Liming Zhu\IEEEauthorrefmark{3}}
\IEEEauthorblockA{\IEEEauthorrefmark{1}School of Computer Science and Engineering, \\University of New South Wales\\
Email: \{shuai.zhang@student., lina.yao@,xiang.zhang3@student.\}unsw.edu.au}
\IEEEauthorblockA{\IEEEauthorrefmark{2}School of Computer Science and Engineering,\\Nanyang Technological University\\
Email: ytay017@e.ntu.edu.sg}
\IEEEauthorblockA{\IEEEauthorrefmark{3}Data61, CSRIO,\\
Email: \{Xiwei.Xu, Liming.Zhu\}@data61.csrio.au}
}


\maketitle

\begin{abstract}

In the past decade, matrix factorization has been extensively researched and has become one of the most popular techniques for personalized recommendations. Nevertheless, the dot product adopted in matrix factorization based recommender models does not satisfy the inequality property, which may limit their expressiveness and lead to sub-optimal solutions. To overcome this problem, we propose a novel recommender technique dubbed as {\em Metric Factorization}. We assume that users and items can be placed in a low dimensional space and their explicit closeness can be measured using Euclidean distance which satisfies the inequality property. To demonstrate its effectiveness, we further designed two variants of metric factorization with one for rating estimation and the other for personalized item ranking. Extensive experiments on a number of real-world datasets show that our approach outperforms existing state-of-the-art by a large margin on both rating prediction and item ranking tasks.

\end{abstract}


%
\IEEEpeerreviewmaketitle

\section{Introduction}
Recommender systems (RS) are among the most ubiquitous and popular components for modern online business. RS aims to identify the most interesting and relevant items (e.g., movies, songs, products, events, articles etc.) from a large corpus and make personalized recommendation for each user. An increasing number of e-commerce and retail companies are utilizing RS to promote their products, enrich shopping potential, boost sales and encourage customer engagement. For customers, RS is an essential filtering tool for overcoming information overload caused by unmanageable amount of resources and selections. As such, RS is of great value to both companies and end users, and a wide spectrum of recommendation models have been developed in the past decade.

RS predicts user's preference on items based on their historical interactions such as ratings, previous purchases and click-through. In general, there mainly exist two categories of recommender tasks: rating prediction and item ranking. The former targets on estimating real-valued rating score (e.g., 1 to 5) of unrated items to which users may likely give based on observed explicit ratings. Many well-known websites such as Netflix, Movielens, IMDB and Douban collect explicit ratings which could be used as an important indicator of user interests and preferences. In some cases where explicit ratings are absent, rank-based recommendation models are more desirable and practical as abundant implicit feedback (e.g., click, view history, purchase log etc.) can be used for generating personalized ranked lists to accomplish the recommendation goal. In this study, we will investigate the usefulness of metric factorization on both scenarios.



RS has been an active research area for both practitioners and researchers. Since the Netflix contest, many recommender systems resort to matrix factorization (MF) due to its superior performance over non-personalized and memory based approaches. MF factorizes the large user-item interaction matrix into a product of two low-dimensional matrices. MF assumes that users and items can be represented by latent factors in a low dimensional space, and models their similarities with dot product.  In recent years, many variants based on matrix factorization such as, SVD~\cite{Koren:2008:FMN:1401890.1401944}, Bayesian personalized ranking~\cite{Rendle:2009:BBP:1795114.1795167}, weighted regularized matrix factorization~\cite{Hu:2008:CFI:1510528.1511352}, probabilistic matrix factorization~\cite{Salakhutdinov:2007:PMF:2981562.2981720} have been developed. To some extent, MF has become a dominant pillar for building up recommender systems despite the fact that its performance may be hindered by the simple choice of interaction method: dot product~\cite{he2017neural,Shi:2014:CFB:2620784.2556270}. It is known that dot product does not satisfy the triangle inequality~\cite{Ram:2012:MIS:2339530.2339677} which is defined as: ``the distance between two points cannot be larger than the sum of their distances from a third point"~\cite{tversky1982similarity}. This weakness limits the expressiveness of matrix factorization and may result in sub-optimal solutions. Detailed explanations are provided in Section 3. Moreover, empirical studies also show that matrix factorization is prone to overfitting for large size of latent factors, which substantially restrict the model flexibility and capability.

To mitigate the above mentioned problem, we propose a novel technique: \textbf{Metric Factorization}, which is applicable to both rating prediction and item ranking. The rational is to replace the dot product of MF with Euclidean distance, and consider the recommendation problem from a position and distance perspective. It factorizes the interaction matrix into user and item dense embedding, and makes recommendations based on their estimated distances. The main contributions of this work are summarized as follows:

\begin{itemize}
    \item We proposed a novel technique, metric factorization,  in which users and items are represented as points in a multi-dimensional coordinate system. Their distances are measured with the Euclidean distance. Recommendations are made based on user and item closeness.


    \item We specify two variants of metric factorization to solve the two classic recommendation tasks: rating prediction and item ranking. Our model can effectively learn the positions of users and items in both settings. Source code of metric factorization is available at Github\footnote{https://github.com/cheungdaven/metricfactorization}.


    \item Extensive experiments on a wide spectrum of large scale datasets demonstrate that our model outperforms the state-of-the-art models by a large margin in terms of both rating estimation and item ranking tasks.
\end{itemize}

The rest of this paper is structured as follows. Section 2 conducts a brief review on studies that are highly relevant to our model. Section 3 describes the research problem.  Section 4 introduces the basic concept of factorization machine. Section 5 to 7 introduce how to perform rating prediction and item ranking with metric factorization respectively. Section 8 demonstrates the model efficacy based on extensive experiments on real-life datasets, and analyzes the model parameters and Section 9 concludes this paper.

\section{Related Work}

Matrix factorization is one of the most effective methods for item recommendation. The first version of matrix factorization for recommender task is designed by Simon Funk\footnote{http://sifter.org/simon/journal/20061211.html} in Netflix contest for rating prediction. Later studies improved MF and proposed many variants. For example, Koren et al.~\cite{Koren:2009:MFT:1608565.1608614} introduced user and item biases to model user and item specific features. Salakhutdinov et al.~\cite{Salakhutdinov:2007:PMF:2981562.2981720,salakhutdinov2008bayesian} interpret MF into a probabilistic graphical model to alleviate the sparsity and in-balance in real-world datasets. Matrix factorization can also be generalized to solve personalized item ranking problem. Two classic top-N recommendation models based on matrix factorization are: Bayesian Personalized Ranking (BPR)~\cite{Rendle:2009:BBP:1795114.1795167} and Weighted Regularized matrix factorization (WRMF)~\cite{Hu:2008:CFI:1510528.1511352}. BPR tackles the item ranking problem from a Bayesian perspective, and it aims to push unobserved user item pairs away from observed user item pairs. WRMF is another effective approach for item ranking. WRMF uses implicit binary feedback as training instances and considers all entries (including unobserved entries) of the user item interaction matrix. WRMF also has a confidence value used to control the weight of negative and positive entries. Incorporating side information into MF is also viable, nonetheless, this work mainly focuses on overcoming the limitation of dot product and we leave the side information modelling with metric factorization as a future work.



As aforementioned, despite the success of MF, it capabilities are constrained by dot product. To this end, several attempts are made to overcome this problem. One approach is to introduce non-linearity to matrix factorization~\cite{zhang2017deep}. Dziugaite et al.~\cite{DBLP:journals/corr/DziugaiteR15} proposed a neural network generalization of matrix factorization by modeling the user-item relationships with non-linear neural networks. The basic idea is to apply non-linear activation functions over the element-wise product of user and item latent factors. He et al.~\cite{he2017neural} followed this idea and proposed the neural collaborative filtering (NeuMF) model for personalized ranking task. NeuMF consists of a generalized MF and a multilayer perceptron. It treats the one-class collaborative filtering problem as a classification task and optimizes the network with a cross-entropy like loss.  Wu et al.~\cite{Wu:2016:CDA:2835776.2835837} proposing using denoising autoencoder to introduce non-linearity to interaction modelling. However, neural networks require more efforts in hyper-parameters tuning, and these settings are usually not adoptable between different datasets.

Another promising attempt is to directly adopt a metric which satisfies the axiom of triangle inequality. Collaborative metric learning (CML)~\cite{Hsieh:2017:CML:3038912.3052639} is such a method which generalizes metric learning to collaborative filtering. CML follows the idea of the largest margin nearest neighbour algorithm (LMNN)~\cite{Weinberger:2009:DML:1577069.1577078}. LMNN aims to estimate a linear transformation to formulate a distance metric that minimizes the expected kNN classification errors.  LMNN consists of two critical operations: \textit{pull} and \textit{push}. The pull operation acts to pull instances in the same class closer, while the push operation acts to push different labeled instances apart. Strictly, CML does not learn the transformation function but the user vectors and item vectors. It only has a push term, which means that CML can only push items that the user dislikes away while does not provide a strategy to pull items the user likes closer. Hsieh et al.~\cite{Hsieh:2017:CML:3038912.3052639} mentioned that the loss function of CML would pull positive items when encountering imposters. Nevertheless, this kind of indirect pull force is too weak compared with that in LMNN, which might not lead to optimal solutions. For example, CML may push possible recommendation candidates too far away.

One work we would like to mention is \cite{plant2014metric}. Although it has the same name (metric factorization) as ours, the underlying principle, techniques, and the tasks of this model and ours are totally different.




\section{Preliminaries}
In this section, we first define the research problem and then discuss the limitation of matrix factorization.

\subsection{Problem Formulation}



Suppose we have $M$ users and $N$ items, the rating/interaction matrix is represented by $R \in \mathcal{R}^{M \times N}$. Let $u$ and $i$ denote user and item respectively. $R_{ui}$ indicates the preference of user $u$ to item $i$. In most cases, most of the entries in $R$ are unknown, and the aim of recommender system is to predict preference scores for these unseen entries. The value of the $R$ can be explicit ratings such as rating score with range [1-5] or implicit feedback. Explicit ratings can be utilized for rating estimation for unknown entries, while implicit feedback can be used for generating personalized ranked recommendation lists (also known as one-class recommendations or top-n recommendation~\cite{Pan:2008:OCF:1510528.1511402}). Implicit feedback such as view history, browsing log, click count and purchasing record are more abundant and easier to acquire than explicit feedback, but explicit ratings still play a critical role in modern recommender systems. In this work, we will investigate both of them and explore/evaluate them individually. Here, the definition of implicit feedback is as follows:
\begin{equation}
R_{ui}=
\begin{cases}
1, &\mbox{if interaction $<u, i>$ exists}\\
0, &\mbox{otherwise}
\end{cases}
\end{equation}
The interaction can flexibly represent any kind of implicit feedback. $1$ indicates that the user item pair $(u,i)$ is observed. $0$ does not necessarily mean the user dislikes the item, it can also indicate the user does not realize the existence of item $i$. We treat items $i$ as a positive sample for user $u$ if $R_{ui}=1$, and as a negative sample if $R_{ui}=0$.

For a clear presentation, Table \ref{notation} summarizes the notations and denotations used in this paper.

\begin{table}[]
\centering
\caption{Notations and Denotations}
\label{notation}
\begin{tabular}{c|l}
\toprule
Notations & \multicolumn{1}{c}{Descriptions} \\
\midrule
$R$, $M$, $N$ & Rating/interaction matrix, number of users and items.  \\
$Y$,$\Hat{Y}$ & Distance matrix and predicted distance matrix.  \\
$P$, $Q$, $k$ &    User/item positions and its dimensionality.  \\
$b_u$, $b_i$, $\mu$, $\tau$ &    User, item and global biases. scale factor for global bias. \\
$c_{ui}$,  $\alpha$   &   Confidence Value and its Confidence level.\\
$a$, $b$,  &   Distance scale factors.    \\
$l$, $\lambda$, $\eta$  & clip value, biases regularization rate, learning rate. \\
\bottomrule
\end{tabular}
\vspace{-3mm}
\end{table}
\subsection{Limitation of Matrix Factorization}

\begin{figure}[h]
\includegraphics[width=0.48\textwidth]{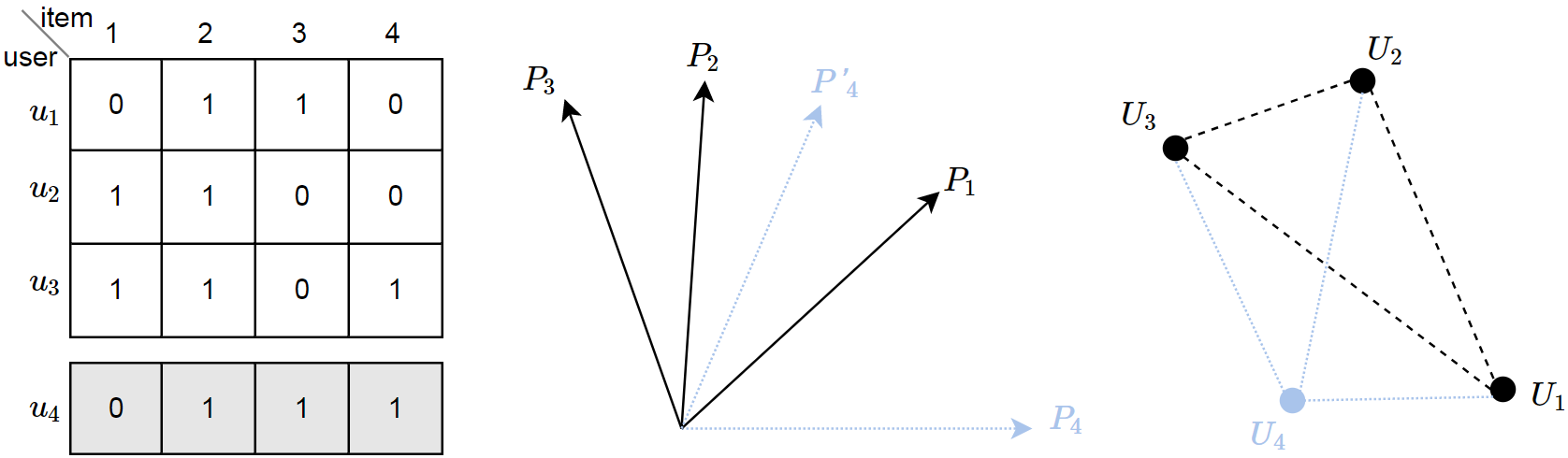}
\caption{An example for explaining the disadvantages of Dot Product; Left: interaction matrix; Middle: user latent factors; Right: positions and distances with metric factorization. All in 2-D dimension.}
\label{innerp}
\end{figure}

Leaving aside the success of matrix factorization for recommendation tasks, it is not without flaws. One major problem is that dot product restricts the expressiveness of matrix factorization~\cite{He:2017:NCF:3038912.3052569}. Dot product cares about the magnitudes and angles of two vectors. To some extent, it measures the similarities rather than distances of the two vectors in terms of both magnitude and angle\footnote{In fact, dot product boils down to cosine similarity when ignoring the magnitude}. Here we extend an example from~\cite{He:2017:NCF:3038912.3052569}. As shown in Figure \ref{innerp}, we attempt to use the user latent factors ($P_1, P_2, P_3, P_4$) to model the similarities of users and compare it with the user similarity computed by Jaccard similarity ($s$) from the interaction matrix. Assuming that we have three users: $u_1$, $u_2$ and $u_3$, the similarities based on interaction matrix are: $s_{23} > s_{12} > s_{13}$, then the corresponding user latent factors $P_1$, $P_2$ and $P_3$ can be positioned like Figure \ref{innerp} (middle). Suppose we have another user $u_4$, and $s_{41} > s_{43} > s_{42}$, in this case, if we make $P_4$ and $P_1$ similar, the constraint $s_{43} > s_{42}$ will not be satisfied no matter how we define $P_4$ in the same space. However, if we treat users as points in the same space, and make similar users closer to each other (that is, to make $\mathcal{D}_{23} < \mathcal{D}_{12} < \mathcal{D}_{13}$, and $\mathcal{D}_{41} < \mathcal{D}_{43} < \mathcal{D}_{42}$, $\mathcal{D}$ denotes distance), we can easily satisfy the aforementioned two constraints.

\begin{figure*}[t]
\centering
\includegraphics[width=0.95\textwidth]{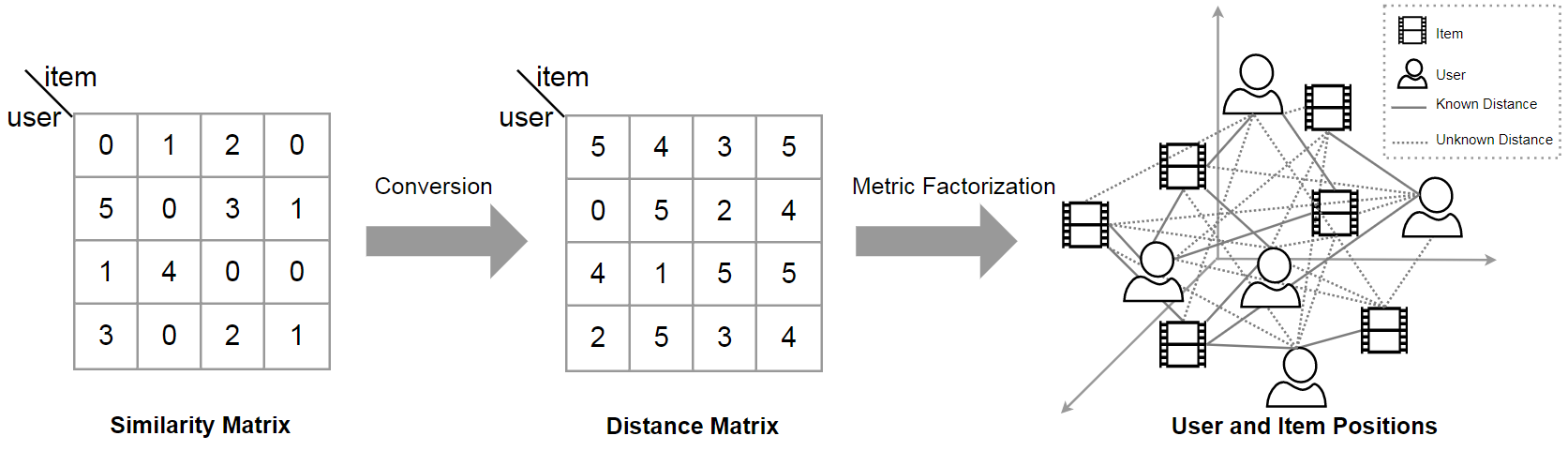}
\caption{A simplified illustration of Metric Factorization. It mainly has two steps: First, convert similarity matrix (rating matrix or interaction matrix) to distance matrix via equation \ref{conversion}. Second, factorize the distance matrix into user and item positions in a low-dimensional space. In the last graph, solid line represents known distances and dash line indicates unknown distances. Once the positions of users \& items are learned, unknown distances can be easily derived for making recommendation.}

\label{conversion}
\end{figure*}

\section{Introduction to Metric Factorization}
To tackle the disadvantage of matrix factorization and tap into the power of positions and distance, we propose a novel recommendation model namely, metric factorization. Unlike matrix factorization which treats users and items as vectors in a new space, metric factorization considers the users and items as points in a low-dimensional coordinate system and their relationships are represented by Euclidean distances. As its name indicates, metric factorization factorizes the distance between users and items to their positions.

Matrix factorization learns user and item latent factors via factorizing the interaction matrix which indicates the similarities between users and items. Nonetheless, to learn the user and item positions, we cannot directly utilize the rating/interaction matrix $R$ since distance and similarity are two opposite concepts. Thus, we need to firstly convert the $R$ into distance matrix $Y$. To this end, we adopt a simple but effective principle:
\begin{equation}
    \textbf{Distance}(u, i) = \textbf{Max Similarity} - \textbf{Similarity}(u,i)
    \label{distance}
\end{equation}
With this equation, highest similarity will become zero which indicates that the distance between $u$ and $i$ is zero. Through this flip operation, we manage to convert similarity to distance while keeping its distribution properties. This conversion can be applied to both explicit rating scores or binary interaction data by just defining the maximum similarity based on the range of feedback. We will specify it in the following sections.

Distance (or Metric, in mathematics, distances and metrics are synonyms, so we will use these two words interchangeably) is a numerical measurement of how far apart objects are. Distance function should satisfy four critical conditions: non-negativity, identity of indiscernibles, symmetry and triangle inequality.  There are many metrics that are used in different application domains or scenarios such as discrete metric, Euclidean metric, Graph metric, etc. In the Euclidean space $\mathcal{R}^k$, distance between two points are usually measured by Euclidean distance (or $\ell_2$-norm distance). That is:
\begin{equation}
    \mathcal{D}(u,i) = \parallel u-i\parallel_2= \left(\sum_{i=1}^k|u_i - i_i|^2\right)^{1/2}
\end{equation}
Euclidean distance has been widely used in machine learning, sensor networks, geometry, etc., due to its simple form and useful properties~\cite{dokmanic2015euclidean}. In practice, squared distance is usually adopted to avoid the trouble of computing the square root.

In metric factorization, we will utilize the Euclidean distance to measure the distance between users and items in a low-dimensional coordinate system.  Let $k$ denote the dimensionality of the space. For each user and item, we assign a vector: $P_u \in \mathcal{R}^k$ and $Q_i \in \mathcal{R}^k$, to determine its position in this multidimensional space. After that, we measure the distance between user and item with Euclidean Distance:
\begin{equation}
    \mathcal{D}(u,i) = \parallel P_u-Q_i\parallel_2^2
    \label{dis}
\end{equation}
In real-world recommender applications, only some entries of the distance matrix are known. The goal of metric factorization is to learn the user and item positions given the partially observed distance matrix:
\begin{equation}
    f(P_u, Q_i|Y)
\end{equation}
The function $f$ varies with the task of recommendation (rating prediction and item ranking). We will detail the formulation with regard to different tasks in the following sections.

So far, we are aware that matrix factorization and metric factorization are very much alike. Although both of them factorize a big and sparse matrix into two more manageable, compact matrices, this factorization operation and the learned matrices have totally different physical interpretations. Nonetheless, we can associate the $P_u$ and $Q_i$ of metric factorization with user's preferences and item latent features in the same way as matrix factorization.




Figure \ref{conversion} simply illustrates the procedure of metric factorization (for rating prediction). First, we get the distance matrix from the similarity matrix with Equation \ref{distance}, then we factorize the distance matrix and learn the positions of users and items. After attaining the positions, if required, we can easily recover every entry of the similarity matrix and make recommendations.


\section{Metric Factorization for Rating Prediction}
As aforementioned, metric factorization can be applied to both rating prediction and item ranking. In this section, we will introduce how to perform rating prediction with metric factorization.

\subsection{Basic Model}
For rating prediction, it is enough and more efficient to just consider observed interaction data.  Let $\mathcal{K}$ denote the set of observed rating data.

First, we convert the rating matrix $R$ into distance metrics with the following equation:
\begin{equation}
    Y_{ui} = R^{max} - R_{ui}
\end{equation}
Where $R^{max}$ is the possible maximum rating score. Note that we use explicit ratings here, thus, if $R^{max}=5$, the rating score of $3$, then the distance $Y_{ui} = 5 -3 =2$.

Afterwards, we need to choose a loss function. Traditional pairwise loss used in metric learning is not suitable for rating prediction as we need to recover the precise rating value instead of a ranked list. Therefore, we adopt the widely used squared loss and we learn the user and item positions by minimizing the following loss:
\begin{equation}
    \mathcal{L} (P, Q) = \sum_{(u,i) \in \mathcal{K}} (Y_{ui} - \parallel P_u-Q_i\parallel_2^2 )^2
    \label{basicmodel}
\end{equation}
This model is closely related to basic matrix factorization,  the only difference is that we replace the dot product with Euclidean distance.



\subsection{Adding Biases and Confidences}
\subsubsection{Adding Biases}
The above basic model only considers the user and item interactions. However, individual effects of users or items also matter. For example, some items tend to receive higher ratings; some users tend to give low rating scores. Same as biased matrix factorization~\cite{Koren:2009:MFT:1608565.1608614}, we also introduce the global, user and item bias terms to metric factorization. Thus, the distance between user and item is formulated as follows:
\begin{equation}
    \hat{Y} = \parallel P_u-Q_i\parallel_2^2 + b_u + b_i + \mu
\end{equation}
Here, $\hat{Y}$ denotes the predicted distances. $b_u$ is the bias term of user $u$; $b_i$ is the bias term of item $i$; $\mu$ is the global bias which equals to the mean distance constructed from training data. Usually, we can add a hyper parameter $\tau$ to scale $\mu$ as the mean distance cannot always reflect the real global bias.

\subsubsection{Confidence Mechanism}
Another important aspect we would like to consider is the reliability and stability of the rating data. Most rating prediction algorithms ignore the noise of ratings and assume that all ratings can be treated as ground truth. Nonetheless, not all observed ratings deserve the same weight~\cite{Koren:2009:MFT:1608565.1608614}. For example, some users might give two different scores when being asked to rate the same item twice at different time~\cite{Jones:2011:CIR:2052138.2052389}. Former studies~\cite{Jones:2011:CIR:2052138.2052389,Amatriain:2009:ILI:1611644.1611670} suggest that extreme ratings (e.g., 1 and 5) are more reliable than moderate ratings (e.g, 2, 3 and 4). To alleviate this issue, we propose adding a confidence value $c_{ui}$ for each rating score and revise the loss function to:
\begin{equation}
    \mathcal{L} = \sum_{(u,i) \in \mathcal{K}} c_{ui} (Y_{ui} - ( \parallel P_u-Q_i\parallel_2^2 + b_u + b_i + \mu ))
    \label{confidence}
\end{equation}
Note that the confidence $c_{ui}$ can represent many factors such as reliability, implicit feedback by specifying the confidence definition. Inspired by the conclusion of~\cite{Jones:2011:CIR:2052138.2052389},  we design a novel confidence mechanism to assign higher confidence values to more reliable ratings.
\begin{equation}
    c_{ui} = 1 + \alpha \cdot g(R_{ui} - \frac{R^{max}}{2})
\end{equation}
Where $g(\cdot)$ can be absolute, square, or even logarithm functions. $\alpha$ (confidence level) is a hyper-parameter to control the magnitude of the confidence. This confidence mechanism ensures extreme ratings to get higher confidence. It is flexible to replace this confidence strategy with other advanced ones.

\subsection{Regularization}
Training on the observed ratings can easily lead to overfitting. Traditional matrix factorization usually adopt $\ell_2$-norm regularization on latent factors and biases to avoid over-complex solutions.
For the user and item biases, we apply the widely used $\ell_2$ norm same as \cite{Koren:2009:MFT:1608565.1608614}.

\subsubsection{Norm Clipping}
For $P$ and $Q$,  $\ell_2$ norm regularization is undesirable as it will push users and items close to the origin. Instead of minimizing the $\ell_2$ regularization, we relax the constraints to the Euclidean ball and perform $\ell_2$ norm clipping after each updates.
\begin{equation}
     \parallel P_u \parallel_2 \leq l , \parallel Q_i \parallel_2 \leq l
\end{equation}
Where $l$ control the size of the Euclidean ball. These two constraints work as regularization to restrict the value of $U_u$ and $V_i$ in $\ell_2$-norm unit ball so that the data points will not spread too widely~\cite{friedman2001elements,Hsieh:2017:CML:3038912.3052639}. This operation is generally performed when updating the parameters in every iteration.

\subsubsection{Dropout}
Dropout is a simple but useful approach for tackling overfitting problem in neural networks~\cite{srivastava2014dropout}. The main idea is to drop some neurons during training stage to avoid co-adaptation between neural units. Here, we propose a similar method and apply it to metric factorization. As Equation (3) indicated, the final distance is the addition of the distance of each dimension. To prevent the co-adaptations among dimensions, we propose randomly dropping the some dimensions and computing the overall distance with the remaining dimensions.
\begin{align}
\parallel P_u-Q_i\parallel_2^2 = |P_{u1} - Q_{i1}|^2 + \underbrace{|P_{u2} - Q_{i2}|^2}_{\textit{drop} } +\nonumber\\ \underbrace{|P_{u3} - Q_{i3}|^2}_{\textit{drop} } +...+|P_{uj} - Q_{ij}|^2+...+\underbrace{|P_{uk} - Q_{ik}|^2}_{\textit{drop} } \nonumber
\end{align}
In the above example, we remove the second, third and $k^{th}$ dimension, so these dimensions will not contribute to the distance prediction in that epoch. The dropped dimensions are subject to change \textit{in each epoch}. Note that this dropout operation is only carried out during the training period.



\section{Metric Factorization for Item Ranking}

\begin{figure}[h]
\centering
\includegraphics[width=0.45\textwidth]{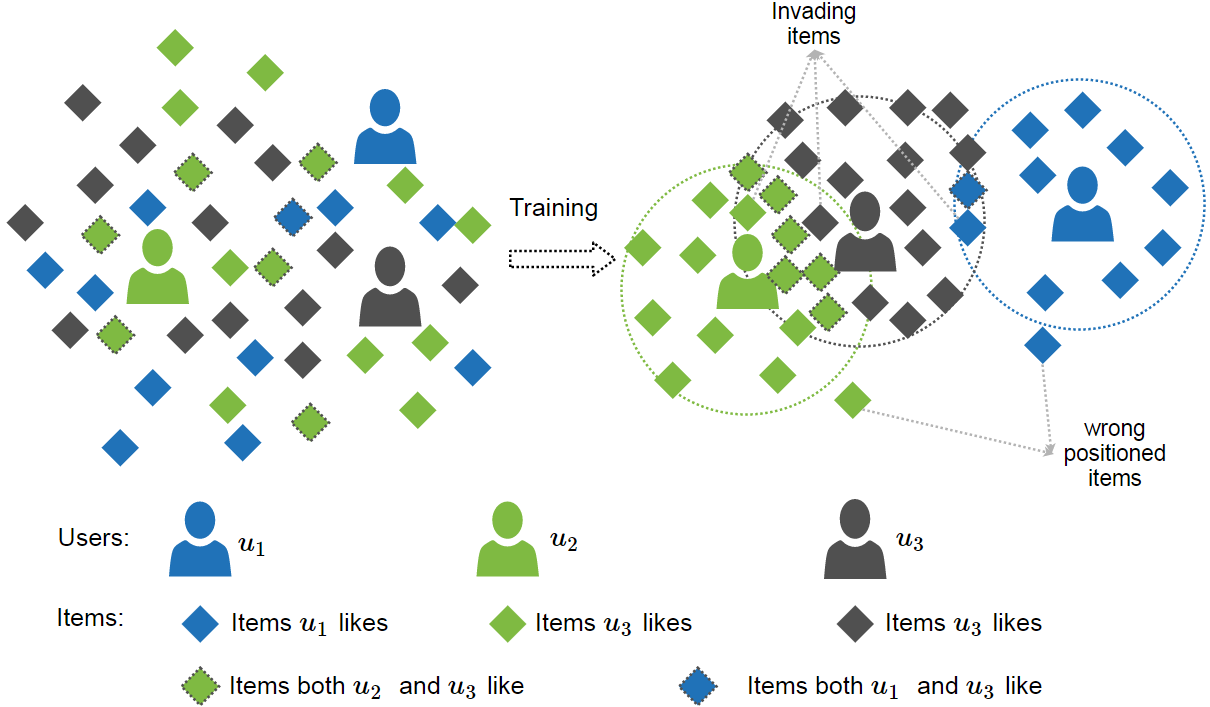}
\caption{Illustration of the proposed approach. Here we give an example which consists of three users and their preferred items. The original positions are randomly initialized. After training, items that user likes will surround her within a finite distance.}
\label{itemranking}
\end{figure}

Item ranking is another important task of recommender systems where only implicit feedback is available. Explicit ratings are not always present in many real-world applications, nonetheless, there is often nontrivial information about the interactions, such as purchase records, listen to tracks, clicks, and even rating itself can also be viewed as a kind of implicit feedback~\cite{koren2008factorization,Zhang:2017:AEH:3077136.3080689}. In reality, the amount of implicit data of some applications far outweighs the quantity of explicit feedback, this makes implicit feedback to be the primary focus in some systems.

First, we need to convert the similarity into distance. Following the principle (2), we conduct the transformation with the following equation:
\begin{equation}
    Y_{ui} = a\cdot(1 - R_{ui}) + z \cdot R_{ui}
\end{equation}
Since $R_{ui}$ equals either $0$ or $1$, the distance $Y_{ui}=a$ if $R_{ui}=0$ and $Y_{ui}=z$ if $R_{ui}=1$. These two hyper-parameters make it very flexible to control the user and item distances. When setting the value of $a$ and $z$, we need to ensure the inequality: $a > z$, to make the distance between user and uninteracted item greater than that between user and interacted item. Usually, we can set $z$ to zero. As $R_{ui}$ equals either ``1" or ``0", all $<\text{user}, \text{positive item}>$ (or $<\text{user}, \text{negative item}>$) has same distance unlike rating prediction whose distances vary with the rating scores.

For ranking task, it is usually beneficial to take the unobserved interactions (negative sample) into consideration. Such as Bayesian personalized ranking, Collaborative metric learning, or weighted matrix factorization ~\cite{Hu:2008:CFI:1510528.1511352}, neural collaborative filtering~\cite{he2017neural}, the former two models are trained in pairwise manner which sample a negative item for each observed interaction. WRMF adopts a pointwise learning approach but also considers negative items. To learn the positions of users and items from implicit data, we employ a pointwise approach and minimize the following weighted square loss.
\begin{equation*}
  \begin{aligned}
    \mathcal{L}(P,Q) =\sum_{(u,i) \in R} c_{ui}( Y_{ui} - \parallel P_u-Q_i\parallel_2^2)^2  \\
s.t. \parallel P_u \parallel_2 \leq l , \parallel Q_i \parallel_2 \leq l
\end{aligned}
\label{cdl}
\end{equation*}
It is similar to equation (7). The major difference is that we consider all unobserved interactions and bias terms are removed.  $c_{ui}$ is also a confidence value, here, it is adopted to model the implicit feedback and is defined as:
\begin{equation}
    c_{ui} = 1 + \alpha w_{ui}
\end{equation}
Where $w_{ui}$ represents the observation of implicit feedback, we can simply treat the counts of implicit actions as the observation, for example, if a user browses the item twice, then $w_{ui}=2$. For large numbers, we can also convert it to log-scale like \cite{Hu:2008:CFI:1510528.1511352}. Since this information is usually absent in publicly available dataset, we set $w$ to $R_{ui}$ (0 or 1). The functionality of hyper-parameter $\alpha$ is the same as that Equation 10. For regularization, we take the same approach including $\ell_2$-norm clipping and dropout as described in Section 5.3.

Figure \ref{itemranking} illustrates an example of how the proposed approach works. Our model can not only force users and their preferred items closer but also push unpreferred items away even though we adopt a pointwise training scheme. Unlike most metric learning based models, which enforce imposters outside of the user territory by some finite margin, the confidence mechanism in this approach provides possibility for negative items to invade user territory, which is beneficial for recommendation task as it can work as a filter to select negative items for recommendation.  Another important characteristic of our model is that it can indirectly cluster users who share a large amount of items together. This property makes it possible to capture the relationships between neighbourhoods which are also important for item recommendation~\cite{1423975}.

\section{Learning of Metric Factorization}
\subsection{Optimization and Recommendation}
We optimize the both the rating estimation and ranking learning model with Adagrad~\cite{duchi2011adaptive}. Adagrad adapts the step size to the parameters based on the frequencies of updates. Thus, it reduces the demand to tune the learning rate. The learning process for metric factorization is shown below.



\begin{algorithm}[h]
  \caption{Training Procedure of Metric Factorization}
  \label{dscfalg}
  \begin{algorithmic}[1]
    \Require $R$, $k$, $P$, $Q$, learning rate $\eta$, $\alpha$, $l$, $a$ and $z$, ($b_u$, $b_i$ , $\tau$, $\lambda$)
    \Procedure{Initialization}{}
        \State initialize $P$, $Q$, (and $b_u$, $b_i$) with normal distribution.
    \EndProcedure
    \Procedure{Model Learning}{}
    \ForAll{pairs $(u, i)$ in current batch}
            \State update $P_u$ and $Q_i$ (and $b_u$, $b_i$) with Adagrad
            \State clip $P_u$ and $Q_i$ by $\ell_2$-norm with $l$
    \EndFor
    \EndProcedure
    \Ensure $P$,$Q$, (and $b_u$, $b_i$)
 \end{algorithmic}
\end{algorithm}

During the testing, we can compute the distance between target user and item positions. As for rating prediction, we reconstruct the ratings using the following rule:
\begin{equation}
    \hat{R}_{ui} = R^{max} - \hat{Y}
_{ui}\end{equation}
For example, if the predicted distance is $\Hat{Y}_{ui}=4$ and $R^{max} =5$, then the predicted rating score is $1$. For item ranking, we directly rank the predicted distance in descending order instead of ascending order as we usually do in matrix factorization. Smaller distance indicates that the user is more likely to be interested in the item. Therefore we will recommend users with items that close to them.

\subsection{Model Complexity} The computation for rating prediction is linear to the number of ratings and it can be trained in $O(|\mathcal{K}|k)$, where $k$ is the dimension of latent factors and $|\mathcal{K}|$ is the number of ratings. For item ranking, since all entries of the interaction matrix are involved, the model complexity is  $O(|R|k)$. As for the computation time (on GPU), it takes about $50s$ to get satisfactory recommendation performance for rating estimation on Movielens 100K. For item ranking, the computation time for each epoch is about $15s$, and it usually takes less than $30$ epochs to achieve a satisfactory result.

\section{Experiments}
In this section, we describe the evaluation settings and show the performance of metric  factorization. As rating prediction and item ranking are usually investigated separately with different evaluation metrics, we evaluate the metric factorization based rating prediction and item ranking models individually. We implement metric factorization with Tensorflow\footnote{https://tensorflow.org} and conduct experiments on a NVIDIA TITAN X Pascal GPU. We empirically evaluate the proposed methodology mainly to answer the following three research questions:
\begin{itemize}
    \item \textbf{RQ1} Does the proposed methodology make more accurate rating estimation than MF and neural network based models?
    \item \textbf{RQ2} Can the proposed approach outperform neural network and metric learning approaches in terms of item ranking?
    \item \textbf{RQ3} How do the hyper-parameters impact the model performances?
\end{itemize}

\subsection{Evaluation for Rating Prediction}

\subsubsection{Datasets Descriptions}
We evaluate the effectiveness of our approach on rating prediction using the following datasets:
\begin{itemize}
    \item \textbf{Goodbooks}. This dataset contains six million ratings for about ten thousand most popular books. It is collected from a real-word book website\footnote{http://fastml.com/goodbooks-10k}.

    \item \textbf{Movielens Datasets}. We use three versions of publicly available Movielens data including Movielens 100K,Movielens 1M and Movielens Latest. Movielens 100K and Movielens 1M are widely adopted benchmark movie data collected by GroupLens research Project. Movielens Latest is a comparatively new dataset which includes historical interactions between 1995 to the year 2016.
\end{itemize}
Table \ref{datasetsrating} summarizes the statistics of these four datasets.

\begin{table}[h]
  \caption{Number of users, items, ratings and corresponding density of four adopted datasets for rating prediction.}
  \label{datasetsrating}
  \centering
  \begin{tabular}{l|ccccc}
    \toprule
    Datasets&\#Items&\#Users&\#Ratings  & Density\\
    \midrule
    Goodbooks & 10K & 53K &6M& 1.119\% \\
    Movielens Latest&9.1K&0.7K&100K&1.644\%\\
    Movielens 100K&1.7K&0.9K&100k&6.305\%\\
    Movielens 1M&3.7K&6K&1M&4.468\%\\

  \bottomrule
\end{tabular}
\vspace{-4mm}
\end{table}

\subsubsection{Evaluation Metrics}
We employ two widely used metrics: Root Mean Squared Error (RMSE) and Mean Average Error (MAE), to assess prediction errors. MAE measures the average magnitude of the errors in a set of predictions, while RMSE tends to disproportionately penalize large errors\cite{Ricci:2010:RSH:1941884}. Since these metrics are commonplace for recommendation evaluation, we omit the details of evaluation metrics for brevity.

\subsubsection{Baselines}
We compare our approach with the following baselines.
\begin{itemize}
    \item \textbf{Average}. This approach predicts ratings based on average historical ratings and has two variants: UserAverage and ItemAverage.


   \item \textbf{SlopOne}~\cite{lemire2005slope}.  It is an efficient online rating-based collaborative filtering approach.

    \item \textbf{BPMF}~\cite{salakhutdinov2008bayesian}.  This model considers the matrix factorization from a probabilistic perspective and provides a fully Bayesian treatment by solving it with Markov chain Monte Carlo.

    \item \textbf{BiasedSVD}. This model introduces the bias to both user and items and equips basic matrix factorization to model the user/item specific characteristics.

    \item \textbf{NRR}~\cite{Li:2017:NRR:3077136.3080822}. NRR is a neural rating regression model which captures the user and item interactive relationships with neural network regression.
    \item \textbf{NNMF}~\cite{dziugaite2015neural}. This model replaces the simple dot product of matrix factorization with multi-layered neural network and decomposes it into two latent vectors for both users and items.

\end{itemize}

\begin{table*}[h]
\centering
\caption{
Comparison between our approach and baselines in terms of RMSE and MAE. \\
Best performance is in boldface and second best is underlined.}
\label{resrating}
\begin{tabular}{c|ccccc}
\toprule
\midrule
  Datasets & Goodbooks & Movielens Latest & Movielens 100K  & Movielens 1M \\
\midrule
Models & \multicolumn{4}{c}{\textbf{Root Mean Squared Error }}   \\
\midrule
UserAverage &   $ 0.893 \pm 0.001$      &  $ 0.967 \pm 0.005$  & $1.047 \pm 0.004$&     $1.036 \pm 0.002$  \\
ItemAverage &   $ 0.953 \pm 0.002$        & $ 0.991 \pm 0.008$     & $ 1.035 \pm 0.005$  &  $0.978 \pm 0.001$     \\
SlopOne &    $0.825 \pm 0.001$    & $0.907 \pm 0.007$  &$0.935 \pm 0.001$ &   $0.899 \pm 0.002$   \\
BPMF       &   $0.829 \pm 0.001$     &   $0.932 \pm 0.001$    &        $0.927 \pm 0.003$   &  $0.867 \pm 0.002$       \\


BiasedSVD &  $0.846 \pm 0.001$       &    $\underline{0.868} \pm 0.005$   &       $0.911 \pm 0.003$   &  $0.847 \pm 0.002$       \\

NRR       &      $0.822 \pm 0.001$       &   $0.876 \pm 0.002$    &       $0.909 \pm 0.003$  & $0.875 \pm 0.002$       \\
NNMF      &   $\underline{0.821} \pm 0.001$     &  $0.871 \pm 0.006$   &     $\underline{0.903} \pm 0.002$     &    $\underline{0.843}\pm 0.002$     \\
\textbf{MetricF}   &   $\textbf{0.803} \pm 0.001$    &  $\textbf{0.853} \pm 0.006$    &        $\textbf{0.891} \pm 0.005$   &   $\textbf{0.836} \pm 0.001$   \\
\midrule
Models & \multicolumn{4}{c}{\textbf{Mean Average Error}}   \\
\midrule
UserAverage &     $ 0.700 \pm 0.001$    &  $ 0.753 \pm 0.002$  &$0.839 \pm 0.004$   &    $0.830 \pm 0.002$    \\
ItemAverage &   $ 0.762 \pm 0.002$       &   $ 0.769 \pm 0.004$     & $ 0.825 \pm 0.005$  & $0.782 \pm 0.001$       \\
SlopOne &  $0.638 \pm 0.001$      & $0.693 \pm 0.007$  &$0.737 \pm 0.001$ &  $0.710 \pm 0.002$    \\
BPMF       &    $0.639 \pm 0.001$      &    $0.707 \pm 0.001$   &      $0.725 \pm 0.003$   &   $0.678 \pm 0.001$      \\


BiasedSVD &    $0.668 \pm 0.001$    & $\underline{0.666} \pm 0.002$      &      $0.718 \pm 0.002$     &  $\underline{0.665} \pm 0.003$        \\

NRR       &   $\underline{0.631} \pm 0.003$      &  $0.674 \pm 0.005$     &    $0.717 \pm 0.005$     &  $0.691 \pm 0.002$      \\
NNMF      &    $0.637 \pm 0.001$    &   $0.667 \pm 0.004$     &     $\underline{0.709} \pm 0.002$     &      $0.669 \pm 0.001$\\
\textbf{MetricF}   & $\textbf{0.616} \pm 0.001$        &    $\textbf{0.656} \pm 0.004$    &        $\textbf{0.696} \pm 0.005$   &  $\textbf{0.654} \pm 0.001$      \\
\midrule
\bottomrule
\end{tabular}
\end{table*}

\subsubsection{Experimental Setups}
We optimize our model in mini batch mode. For a fair comparison, we randomly split each dataset into a training set and a test set by the ratio of 9:1, and report the average results over five different splits. Hyper-parameters for compared baselines are carefully tuned to achieve their best performances. Our model is not very sensitive to small variations of learning rate. This is mainly due to the adaptive gradient optimizer we adopted as it can eliminate the demand to tune the learning rate. Thus, we set the learning rate to $0.05$ for all datasets. The bias regularization rate $\lambda$ is set to $0.01$ excerpt for Movielens latest ($\lambda = 0.05$). The confidence level $\alpha$ is set to $0.2$ excerpt for Movielens 1M ($\alpha = 0.1$). $g(\cdot)$ is set to square function for Goodbooks, Movielens latest and 1M and it is set to absolute function for Movielens 100k. Other parameters are listed as follows:
\begin{table}[t]
  \caption{Hyper-parameter settings for rating prediction.}
  \label{datasets}
  \begin{tabular}{l|l}
    \toprule
    Dataset&\multicolumn{1}{c}{\textbf{Hyper-parameters}}\\
    \midrule
    Goodbooks& $k=200$, $dropout = 0.10$, $\tau = 0.80$, $l=1.00$ \\
    Movielens Latest&$k=250$,  $dropout = 0.15$, $\tau = 0.80$, $l=0.85$ \\
    Movielens 100k&$k=150$, $dropout = 0.05$, $\tau = 0.90$, $l=1.00$ \\
    Movielens 1M& $k=150$, $dropout = 0.03$, $\tau = 0.50$, $l=1.40$ \\
  \bottomrule
\end{tabular}
\end{table}

\subsubsection{Results and Discussions}
From Table \ref{resrating}, we observed that our model consistently outperforms all compared methods including classic matrix factorization based models as well as neural network based models on all datasets. Specifically, Metric factorization achieves substantial improvements over BPMF and BiasedSVD. Naive approaches UserAverage and ItemAverage perform poorly on all datasets. Recent neural network based model NRR and NNMF are two competitive baselines especially on large scale dataset Goodbooks. However, NRR does not perform well on other three datasets. With this table, we can give a sure answer to \textbf{RQ1}.

\subsection{Evaluation for Item Ranking}
In this section, we evaluate the performance of metric factorization on item ranking tasks.

\subsubsection{Datasets Description}

We also report the experimental results over four real-world datasets.
The datasets are described as follows:
\begin{itemize}

    \item \textbf{Yahoo Research Dataset}. It consists of two datasets: Yahoo movie and Yahoo music. Both of them are provided by Yahoo Research Alliance Webscope program\footnote{http://research.yahoo.com/Academic\_Relations}. The music preference dataset is collected from Yahoo Music services.

    \item \textbf{FilmTrust}. FilmTrust is crawled from a film sharing and rating website by Guo et al.~\cite{guo2013novel}. In this work, we do not use the trust information in this dataset.

    \item \textbf{EachMovie}. EachMovie\footnote{http://www.cs.cmu.edu/~lebanon/IR-lab/data.html} is also a movie dataset which is widely adopted for recommendation evaluation.

\end{itemize}

\begin{table}[t]
\centering
  \caption{Statistics of datasets used for item ranking.}
  \label{datasetsranking}
  \begin{tabular}{l|cccc}
    \toprule
    Datasets&\#Items&\#Users&\#Interact&\%Density\\
    \midrule
    YahooMovie&11.9K&7.6K&221.4K&0.243\%\\
    YahooMusic&1.0K&15.4K&365.7K&2.374\%\\
    FilmTrust&2.1K&1.5K&35.5K&1.137\%\\
    EachMovie&61.3K &1.6K&2.811M & 2.369\% \\
  \bottomrule
\end{tabular}
\vspace{-2mm}
\end{table}
Table \ref{datasetsranking} summarizes the details of the datasets. All of the interactions are binarized by the method introduced in Section 3.1.

\begin{table*}[t]
\centering
\caption{
Performance comparison on item ranking task in terms of MAP, MRR, NDCG, Precision and Recall on four datasets. Best performance is in boldface and second best is underlined.}
\label{results}
\begin{tabular}{c|ccccccc}
\toprule
\midrule
Methods      & MAP  & MRR & NDCG & P@5 & P@10 & R@5 & R@10\\
\midrule

 \multicolumn{8}{c}{\textbf{YahooMovie}} \\
\midrule
POP     & $0.160 \pm 0.001$ &$0.323 \pm 0.002$ & $0.386 \pm 0.002$  & $0.128 \pm 0.001$ &$0.097 \pm 0.002$ & $0.160 \pm 0.003$  & $0.239\pm 0.002$\\
ItemKNN       & $0.177 \pm 0.001$&$0.338\pm 0.002$&$0.395\pm 0.001$&$0.155\pm 0.003$&$0.122\pm 0.002$&$0.180 \pm 0.001$ & $ 0.275 \pm 0.002$\\
BPR &$0.161 \pm 0.001$ &$0.322 \pm 0.001$&$0.387 \pm 0.001$&$0.128 \pm 0.001$&$0.096\pm 0.002$ &$ 0.161 \pm 0.001$ & $ 0.240 \pm 0.001$\\
NeuMF &$0.171 \pm 0.001$&$0.329 \pm 0.006$&$0.402 \pm 0.004$&$0.140 \pm 0.002$&$0.112 \pm 0.001$ & $ 0.168 \pm 0.002$ & $0.256 \pm 0.003$\\
WRMF &$0.209 \pm 0.001$&$0.384 \pm 0.002$&$0.435 \pm 0.004$&$0.176 \pm 0.001$ & $0.135 \pm 0.001$ &  $0.209 \pm 0.001$  &  $0.310 \pm 0.001$  \\
CDAE &$0.224 \pm 0.002$&$0.409 \pm 0.005$ &$0.453 \pm 0.003$ &$0.188 \pm 0.003 $ & $0.143 \pm 0.002 $ & $0.226 \pm 0.003 $& $0.331 \pm 0.004 $ \\
CML-PAIR &$0.223 \pm 0.002$&$0.416 \pm 0.003$&$0.456 \pm 0.003$ & $0.189 \pm 0.004$&$0.147 \pm 0.002$ & $0.224 \pm 0.003$ &  $0.332 \pm 0.002$\\
CML-WARP &$\underline{0.235} \pm 0.002$&$\underline{0.440} \pm 0.005$&$\underline{0.466} \pm 0.001$ & $\underline{0.202} \pm 0.001$& $\underline{0.153} \pm 0.001$ & $\underline{0.237} \pm 0.002$ & $\underline{0.342} \pm 0.002$  \\
MetricF &$\textbf{0.262} \pm 0.001$&$\textbf{0.466} \pm 0.002$&$\textbf{0.486} \pm 0.001$ & $\textbf{0.222} \pm 0.001$ &$\textbf{0.167} \pm  0.001$ &$\textbf{0.262} \pm 0.002 $ & $\textbf{0.374} \pm 0.002$ \\
\midrule
\multicolumn{8}{c}{\textbf{YahooMusic}} \\
\midrule
POP     & $0.100 \pm 0.001$ &$0.237 \pm 0.003$ & $0.334 \pm 0.001$ & $0.088 \pm 0.001$& $0.072 \pm 0.002$& $0.103 \pm 0.002$& $0.166 \pm 0.002$\\
ItemKNN        &$0.132 \pm 0.002$&$0.299 \pm 0.001$&$0.375 \pm 0.002$& $0.115 \pm 0.002$& $0.090 \pm 0.002$& $0.135 \pm 0.003$& $0.207 \pm 0.001$\\
BPR &$0.133 \pm 0.001$&$0.300 \pm 0.002$&$0.373 \pm 0.001$& $0.118 \pm 0.001$& $0.092 \pm 0.001$& $0.141 \pm 0.002$& $0.215 \pm 0.001$\\
NeuMF &$0.141 \pm 0.002$&$0.319 \pm 0.005$&$0.383 \pm 0.003$& $0.130 \pm 0.002$& $0.099 \pm 0.001$& $0.155 \pm 0.001$& $0.232 \pm 0.003$\\
WRMF &$0.144 \pm 0.001$&$0.333 \pm 0.003$&$0.386 \pm 0.001$& $0.130 \pm 0.002$& $0.097 \pm 0.001$& $0.153 \pm 0.002$& $0.226 \pm 0.002$\\
CDAE &$\underline{0.153} \pm 0.003$&$0.348 \pm 0.006$ &$\underline{0.395} \pm 0.004$ & $0.136 \pm 0.002$ & $0.102 \pm 0.001$ & $0.163 \pm 0.002$ & $0.240 \pm 0.003$ \\
CML-PAIR &$0.148 \pm 0.002$&$0.339 \pm 0.003$&$0.390 \pm 0.003$& $0.134 \pm 0.003$& $0.102 \pm 0.002$& $0.159 \pm 0.002$& $0.236 \pm 0.004$\\
CML-WARP &$\underline{0.153} \pm 0.002$&$0.353 \pm 0.005$&$0.393 \pm 0.002$& $\underline{0.140} \pm 0.002$& $\underline{0.103} \pm 0.001$& $\underline{0.171} \pm 0.002$& $\underline{0.249} \pm 0.003$\\
MetricF &$\textbf{0.169} \pm 0.001$&$\textbf{0.375} \pm 0.002$&$\textbf{0.411} \pm 0.001$ & $\textbf{0.152} \pm 0.001$ &$ \textbf{0.114} \pm 0.001$ &$ \textbf{0.183} \pm 0.001$ &$ \textbf{0.268} \pm 0.002$ \\

\midrule
\multicolumn{8}{c}{\textbf{EachMovie}} \\
\midrule
POP & $0.259 \pm 0.001$ &$0.457 \pm 0.001$&$0.522 \pm 0.001$&$0.227 \pm 0.001$&$0.172 \pm 0.002$ &$0.203 \pm 0.001$&$0.267 \pm 0.002$\\
ItemKNN   &$0.407 \pm 0.001$&$0.636 \pm 0.002$&$0.649 \pm 0.001$&$0.364 \pm 0.002$&$0.291 \pm 0.001$&$0.333 \pm 0.002$&$0.469 \pm 0.003$\\
BPR &$0.394 \pm 0.003$&$0.627 \pm 0.002$&$0.641 \pm 0.002$&$0.346 \pm 0.001$&$0.293 \pm 0.001$&$0.313 \pm 0.001$&$0.448 \pm 0.001$\\
NeuMF &$0.414 \pm 0.003$&$0.656 \pm 0.002$&$0.657 \pm 0.003$&$0.378 \pm 0.003$&$0.302 \pm 0.002$&$0.335 \pm 0.003$&$0.475 \pm 0.001$\\
WRMF &$\underline{0.433} \pm 0.001$&$0.679 \pm 0.001$&$0.670 \pm 0.002$&$0.397 \pm 0.002$&$0.314 \pm 0.001$&$0.355 \pm 0.002$&$0.494 \pm 0.001$\\
CDAE &$0.432 \pm 0.003$&$0.678 \pm 0.003$ &$\underline{0.673} \pm 0.002$&$0.394 \pm 0.004$&$0.311 \pm 0.003$&$\underline{0.356} \pm 0.003$&$\underline{0.497} \pm 0.002$\\
CML-PAIR &$0.426 \pm 0.003$&$0.674 \pm 0.002$&$0.668 \pm 0.001$&$\underline{0.399} \pm 0.001$&$\underline{0.315} \pm 0.002$&$0.348 \pm 0.002$&$0.492 \pm 0.001$\\
CML-WARP &$0.419 \pm 0.001$&$\underline{0.683} \pm 0.001$&$0.663 \pm 0.002$&$0.398 \pm 0.001$&$0.312 \pm 0.001$&$0.346 \pm 0.002$&$0.485 \pm 0.003$\\
MetricF &$\textbf{0.454} \pm 0.001$&$\textbf{0.696} \pm 0.002$&$\textbf{0.687} \pm 0.001$&$\textbf{0.416} \pm 0.001$&$\textbf{0.330} \pm 0.001$&$\textbf{0.370} \pm 0.002$&$\textbf{0.515} \pm 0.002$\\

\midrule
\multicolumn{8}{c}{\textbf{FilmTrust}} \\
\midrule
POP     & $0.489 \pm 0.002$ &$0.618 \pm 0.004$ & $0.650 \pm 0.002$ &$0.418 \pm 0.004$ &$0.350 \pm 0.002$ &$0.397 \pm 0.008$ &$0.631 \pm 0.004$\\
ItemKNN   & $0.222 \pm 0.009$ &$0.362 \pm 0.037$ & $0.331 \pm 0.006$ &$0.196 \pm 0.004$  &$0.180 \pm 0.004$  &$0.218 \pm 0.004$  &$0.351 \pm 0.002$\\
BPR &$0.476 \pm 0.004$&$0.600 \pm 0.007$&$0.635 \pm 0.003$ &$0.412 \pm 0.005$ &$0.347 \pm 0.001$ &$0.391 \pm 0.009$ &$0.613 \pm 0.007$\\
NeuMF &$0.483 \pm 0.001$&$0.609 \pm 0.005$&$0.646 \pm 0.003$ &$0.413 \pm 0.003$ &$0.350 \pm 0.002$ &$0.393 \pm 0.004$ &$0.626 \pm 0.007$\\
WRMF &$0.516 \pm 0.002$&$0.648 \pm 0.005$&$0.663 \pm 0.002$ &$0.433 \pm 0.002$ &$0.351 \pm 0.001$ &$0.427 \pm 0.005$ &$0.632 \pm 0.007$\\
CDAE &$0.523 \pm 0.008$&$0.654 \pm 0.010$ &$0.678 \pm 0.008$ &$0.436 \pm 0.004$ &$0.353 \pm 0.003$&$\underline{0.441} \pm 0.006$&$0.647 \pm 0.005$ \\
CML-PAIR &$0.491 \pm 0.002$&$0.637 \pm 0.003$&$0.655 \pm 0.001$&$0.418 \pm 0.002$&$0.337 \pm 0.001$&$0.408 \pm 0.003$&$0.602 \pm 0.003$ \\
CML-WARP &$\underline{0.529} \pm 0.004$&$\underline{0.666} \pm 0.005$&$\underline{0.684} \pm 0.003$&$\underline{0.438} \pm 0.005$ &$\underline{0.359} \pm 0.003$&$\underline{0.441} \pm 0.007$&$\underline{0.653} \pm 0.004$ \\
MetricF &$\textbf{0.549} \pm 0.005$&$\textbf{0.685} \pm 0.006$&$\textbf{0.701} \pm 0.004$&$\textbf{0.453} \pm 0.003$&$\textbf{0.366} \pm 0.002$&$\textbf{0.462} \pm 0.006$&$\textbf{0.672} \pm 0.006$\\
\midrule
\bottomrule

\end{tabular}

\end{table*}

\subsubsection{Evaluation Metrics}
To evaluate the ranking accuracy and quality, we employed five widely used metrics: Recall, Precision, Mean Average Precision (MAP), Mean Reciprocal Rank (MRR) and Normalized Discounted Cumulative Gain (DNCG). In most cases, users only care about the topmost recommended items, so we employ these evaluations at a given cut-off $n$ and denote them as: Precision@n and Recall@n. In practice, making items that interest target users more rank higher will enhance the quality of recommendation lists. MAP, MRR and NDCG are three rank-aware measurements with which higher ranked positive items are prioritized. Thus they are suitable for assessing the quality of ranked lists~\cite{shani2011evaluating}. MRR cares about the single highest-ranked relevant item. NDCG is determined by Discounted Cumulative Gain and Ideal Discounted Cumulative Gain. Detail definitions are omitted for the sake of conciseness.


\subsubsection{Baselines}
We compare our model with the following classic and recent strong baselines:
\begin{itemize}
    \item \textbf{POP}. It is a non-personalized method which generates recommendations according to the item popularity and recommends users with the most popular items.

    \item \textbf{ItemKNN}~\cite{deshpande2004item}. Item-based collaborative filtering method recommends items which are similar to items the user has liked. Here the similarity between items is computed with cosine function.

    \item \textbf{BPR}~\cite{Rendle:2009:BBP:1795114.1795167}. It is a generic pairwise optimization criterion and works even better by integrating MF. The BPR optimization criterion aims to maximize the differences between negative and positive samples.

    \item \textbf{WRMF}~\cite{Hu:2008:CFI:1510528.1511352}. It is an effective ranking model for implicit feedback. It uses dot product to model the user item interaction patterns as well.

    \item \textbf{NeuMF}~\cite{He:2017:NCF:3038912.3052569}. NeuMF combines multi-layered preceptron with generalized matrix factorization and computes the ranking scores with a neural layer instead of dot product. It treats the ranking problem as a classification task and optimizes the model by minimizing the cross entropy loss.

    \item \textbf{CDAE}~\cite{Wu:2016:CDA:2835776.2835837}. CDAE learns user and item distributed representations from interaction matrix with autoencoder. Denoising technique is utilized to improve generalization and prevent overfitting. Here, we use logistic loss to optimize the network as it is reported to achieve the best performances in the original paper.

    \item \textbf{CML}~\cite{Hsieh:2017:CML:3038912.3052639}. CML is competitive baseline for the proposed approach. Here, we train it with two types of loss: pairwise hinge loss and weighted approximate-rank pairwise loss (WAPR). Note that the WARP loss is computational intensive due to the pre-computations for negative samples. Here, we set the negative sampling number to 20 to avoid extended computing time.

\end{itemize}
BPR and WRMF are matrix factorization based models; NeuMF and CDAE are neural network based models; CML-PAIR and CML-WARP are metric learning based models.


\subsubsection{Implementation Details}
We adopt the implementation of  Mymedialite\footnote{http://mymedialite.net/index.html} for POP, ItemKNN, BPR and WRMF. For NeuMF, we use their released source code\footnote{https://github.com/hexiangnan/neural\_collaborative\_filtering}. We implemented all other models with Tensorflow.  We test our model with five random splits by using 80\% of the interaction data as training set and the remaining 20\% as test set and report the average results.  Hyper-parameters are determined with grid search. The dimensions of user and item vectors in our approach and CML are tuned amongst $\{80, 100, 120, 150, 200\}$. Learning rate of the proposed model is set to $0.1$.  The margin of CML-pair and CML-WARP is amongst $\{0.5, 1.0, 1.5, 2.0 \}$. For all datasets, we consider the interaction as the observation of implicit feedback and set $w_{ui} = R_{ui}$ (Here, $R_{ui}$ is constructed from the training set.). Same as rating prediction, we set the learning rate to $0.05$ for all datasets. The dimension $k$ is set to $100$ for EachMovie and $k=200$ for other three datasets. We find that simply setting the distance scale factor $z$ to $0$ is sufficient for the adopted four datasets. Another distance scale factor $a$ is set to: $2.25$ for FilmTrust, $2.0$ for YahooMovie and EachMovie, $0.5$ for YahooMusic. The confidence level $\alpha$ is set to $4$ for FilmTrust and $1$ for other three datasets. Dropout does not influence the ranking performance much on these four datasets, so we do not use dropout in the experiments.

\subsubsection{Results and Discussions}

Results in terms of the MAP, MRR, NDCG, Precion and Recall at cut-off five and ten are shown in Table \ref{results}.  In particular, three key observations can be drawn from the results:

First, the proposed approach consistently achieves the best performances on all datasets. The overall performance gaining over WRMF model is up to $20.8\%$. Although both CML and the proposed approach rely on distance measurement, our model  still outperforms the CML by up to $8.7\%$, which indicates that our model can reflect the distance relationship among users and items more accurately. Second, Euclidean distance based models (Metric factorization and CML) outperform methods that use dot product, neural network models and neighbourhood based models; The improvement suggests that metric based models are more plausible to model user item interaction patterns than other forms. Third, neural network based models CDAE and NeuMF are strong baselines and outperform traditional matrix factorization. Even though NeuMF and BPRMF adopt negative sampling, NeuMF gets more accurate results than BPRMF. CDAE performs better than WRMF in most cases. This suggests the certain effectiveness of introducing non-linearity. However, there is still a huge performance gap between these models and our approach, which also demonstrates the superiority of metric factorization compared to neural based approach.

In words, these observations demonstrate the advantages of considering the recommendation problem from the positions and distances perspective, and the experiments empirically show the efficacy of the method we adopted to learn users' and items' positions.  With these results, we can give a sure answer to research question \textbf{RQ2}.

\subsection{Model Parameter Analysis}
In this section, we will investigate the impact of model parameters to answer \textbf{RQ3}. All the experiments are performed on Movielens Latest (Rating Prediction) and FilmTrust (Item Ranking). Due to limited space, we only report the RMSE for rating prediction and NDCG for item ranking. Matrix factorization model BiasedSVD and WRMF act as the baselines.

\begin{figure}[h]
\begin{minipage}[t]{4.2cm}
\includegraphics[width=4.2cm]{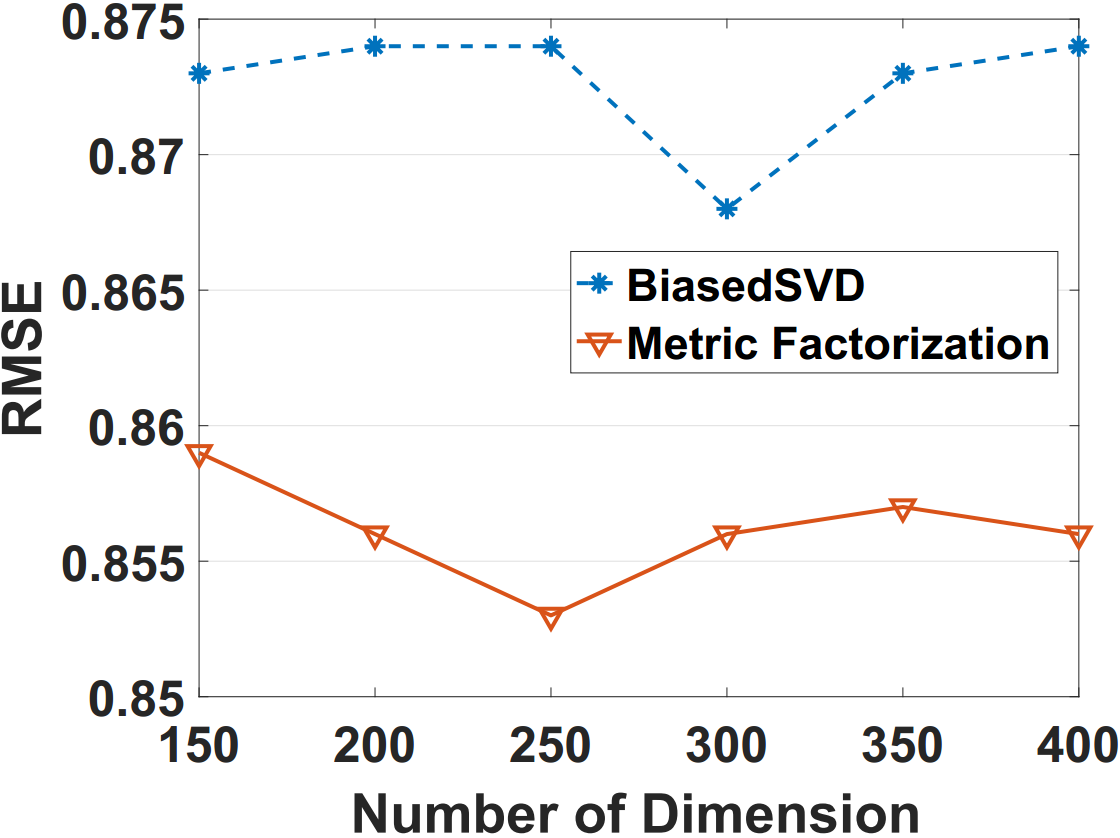}
\centering{(a). Rating Prediction}
\end{minipage}
\begin{minipage}[t]{4.1cm}
\includegraphics[width=4.1cm]{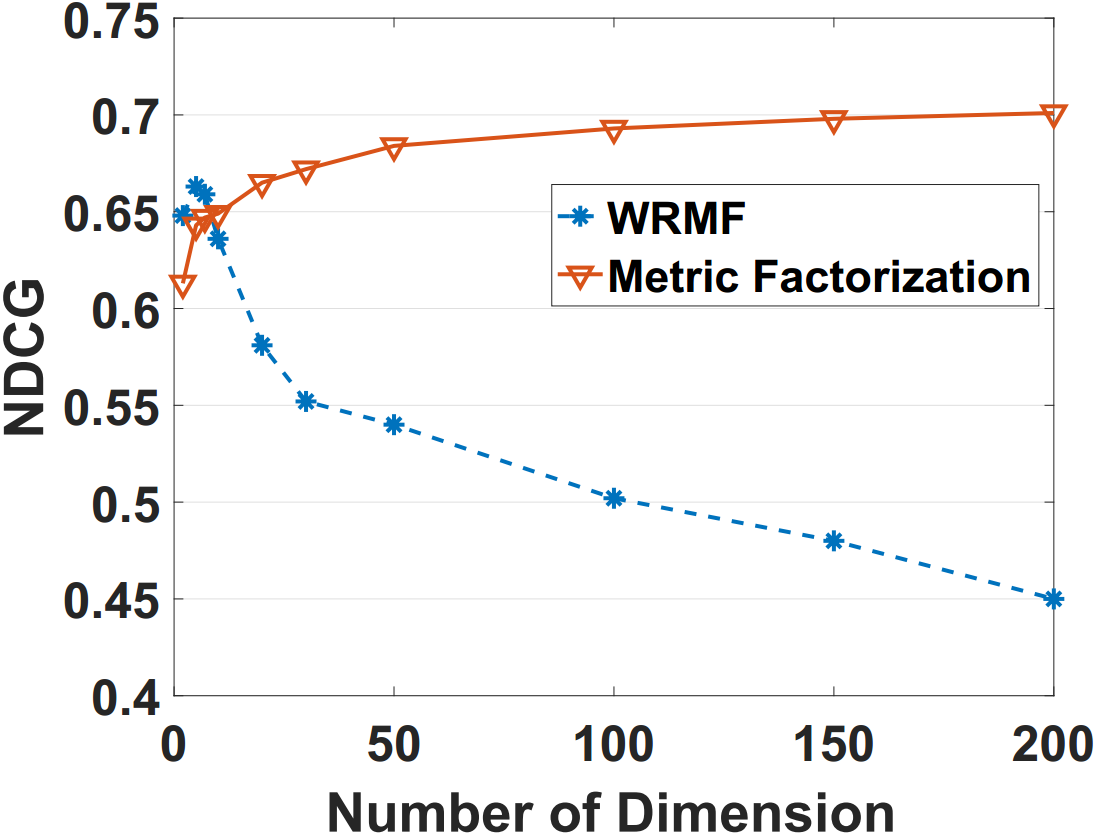}
\centering{(b). Item Ranking}
\end{minipage}
\caption{(a) RMSE with varying Number of Dimension ($k$); (b) NDCG with varying  Number of Dimension ($k$).}
\label{fig:k}
\end{figure}

\subsubsection{Number of Dimension $k$}
Figure \ref{fig:k}(a) shows the varying RMSE on rating prediction for Movielens Latest. We observe that our model outperforms BiasedSVD with same dimension settings, and it takes smaller dimension for metric factorization to reach the best performance. The overfitting issue on rating prediction of both models is not serious. Figure \ref{fig:k}(b) shows the varying NDCG on item ranking for FilmTrust. Clearly, WRMF is easy to overfit with large number of dimensions. Best performance of WRMF is usually achieved with a comparatively small number, and then the performance will decrease rapidly (a similar phenomenon is also observed on other datasets.), which greatly limits the flexibility and expressiveness of the recommendation model. This might be due to the weakness of dot product we explained in Section 3.2.  While our model is less likely to overfit with comparatively large number of dimensions, the performances are usually enhanced by increasing the dimension size properly, which allows the model to capture more factors of user preferences and item characteristics on personalized ranking task.

\begin{figure}[t]
\begin{minipage}{4.2cm}
\includegraphics[width=4.2cm]{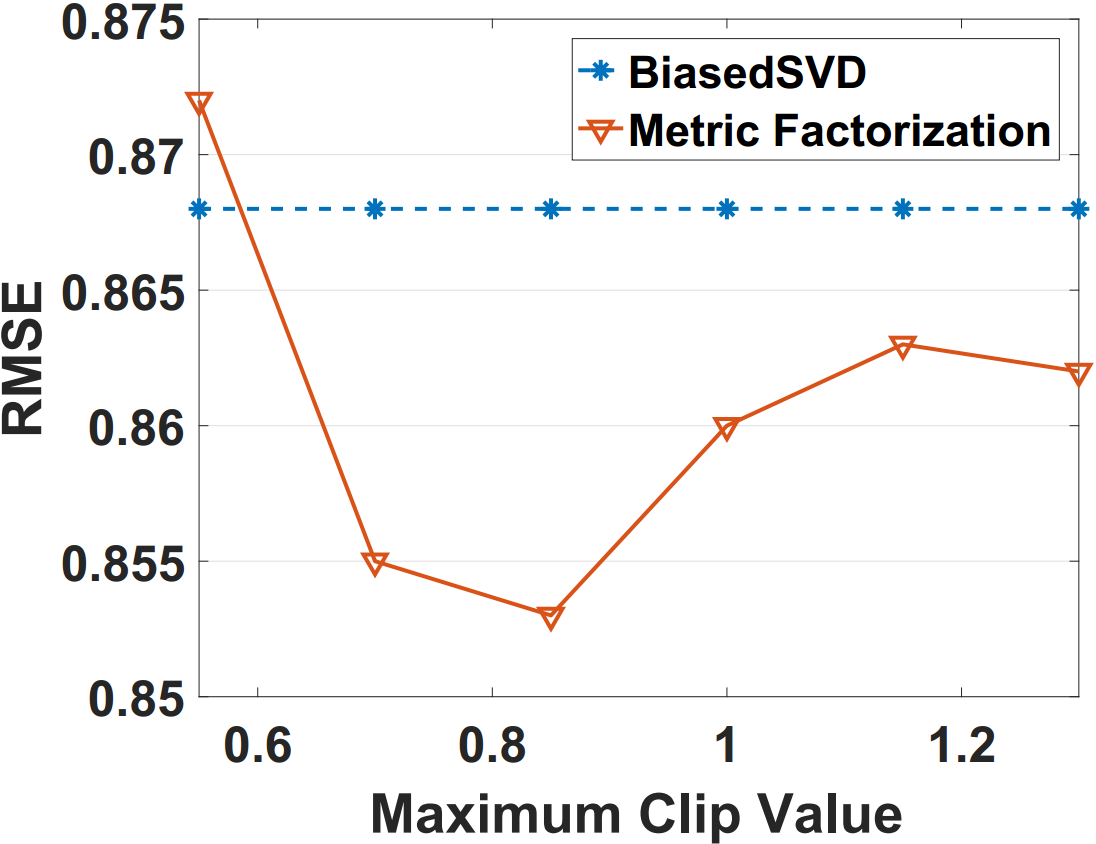}
\centering{(a). Rating Prediction}
\end{minipage}
\begin{minipage}{4.1cm}
\includegraphics[width=4.1cm]{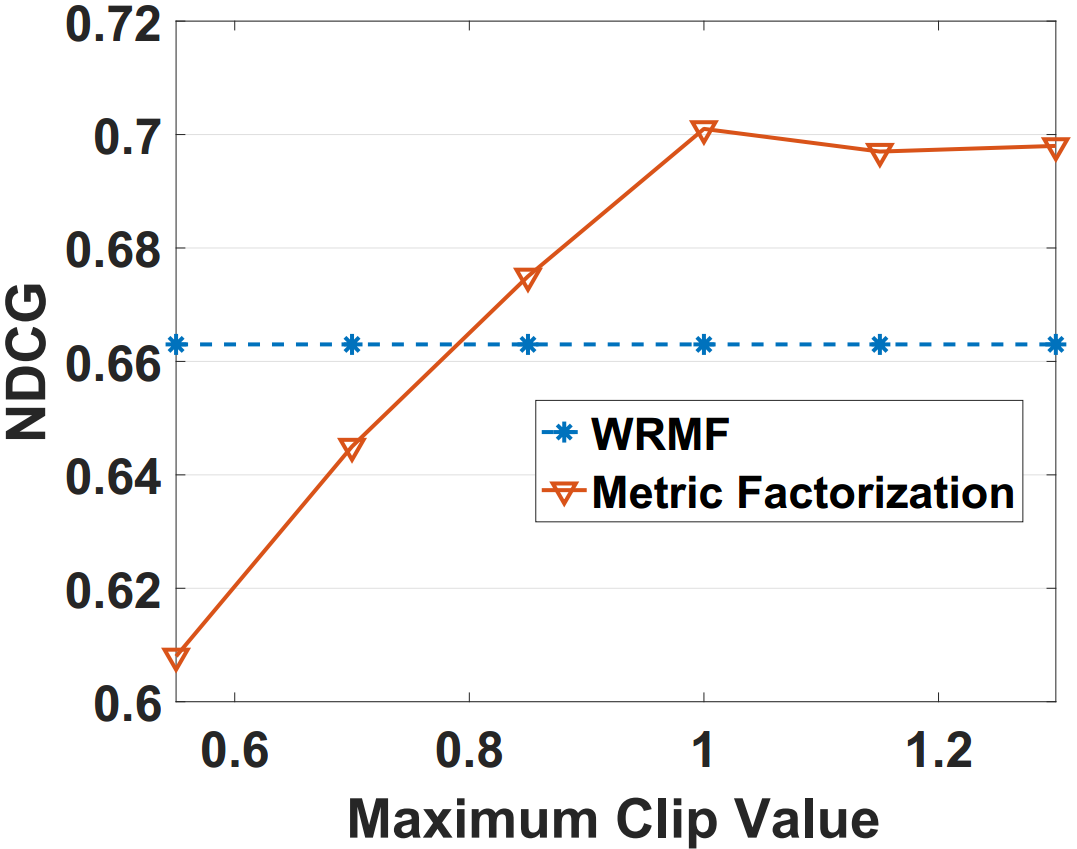}
\centering{(b). Item Ranking}
\end{minipage}
\caption{(a) RMSE with varying maximum clip value; (b) NDCG with varying maximum clip value.}
\label{fig:clip}
\end{figure}

\subsubsection{Maximum Clip Value $l$}
Figure \ref{fig:clip} shows the impacts of maximum clip value $l$ on model performance. $l$ is a critical model parameters as it controls the scope of user and item positions. Intuitively, ideal clip value might be determined by the range of feedback value (e.g., ratings, binary implicit feedback). Since the implicit feedback in our work is set to $0$ or $1$, setting $l$ to $1$ is enough in most cases. While for rating prediction, the rating score range could be diverse, so $l$ should be carefully tuned.

\begin{figure}[h]
\begin{center}
\begin{minipage}[t]{4.25cm}
\includegraphics[width=4.25cm]{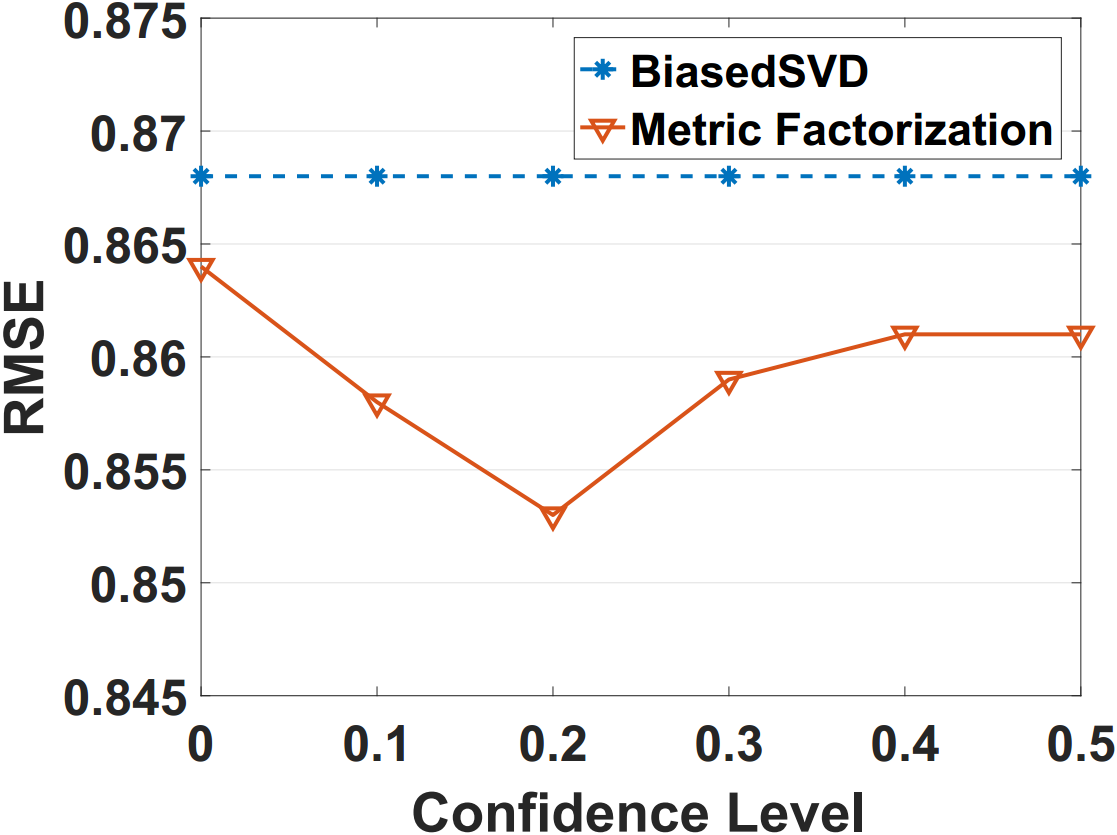}
\centering{(a). Rating Prediction}
\end{minipage}
\begin{minipage}[t]{4.1cm}
\includegraphics[width=4.1cm]{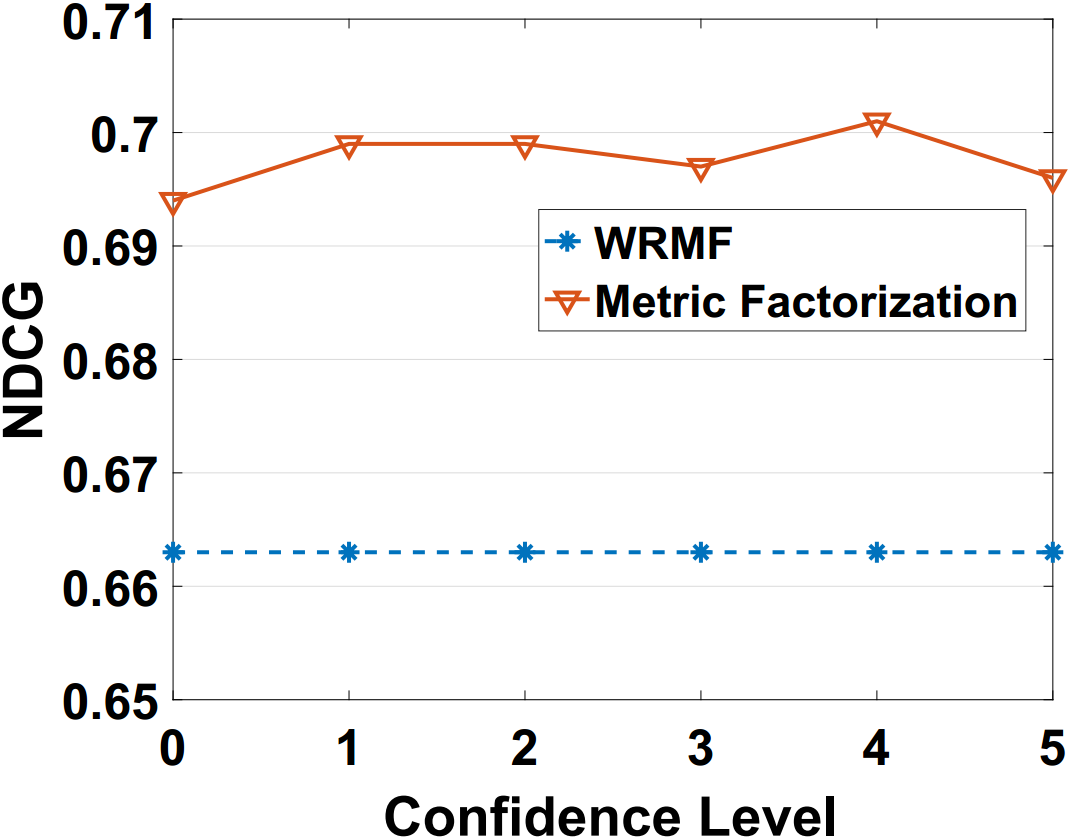}
\centering{(b). Item Ranking}
\end{minipage}
\caption{(a) RMSE with confidence level $\alpha$; (b) NDCG with confidence level $\alpha$.}
\label{fig:conf}
\end{center}
\end{figure}

\subsubsection{Confidence Level $\alpha$}
In this experiment, we mainly investigate the impact of hyper-parameter $\alpha$. Setting $\alpha=0$ means we do not employ the confidence mechanism.  It reflects the model's confidence on its prediction. In rating prediction, $\alpha$ works with function $g(\cdot)$ to reduce the potential noise in rating data and increase the model robustness. In item ranking, $\alpha$ scales the implicit feedback to differentiate between positive samples and negative samples. As shown in Figure \ref{fig:conf}, the confidence mechanism does help improve the model efficacy. One practical guideline for choosing $\alpha$ is to determine it based on how much confidence you have about your estimation.


\begin{figure}[!h]
\begin{center}
\begin{minipage}[t]{4.2cm}
\includegraphics[width=4.2cm]{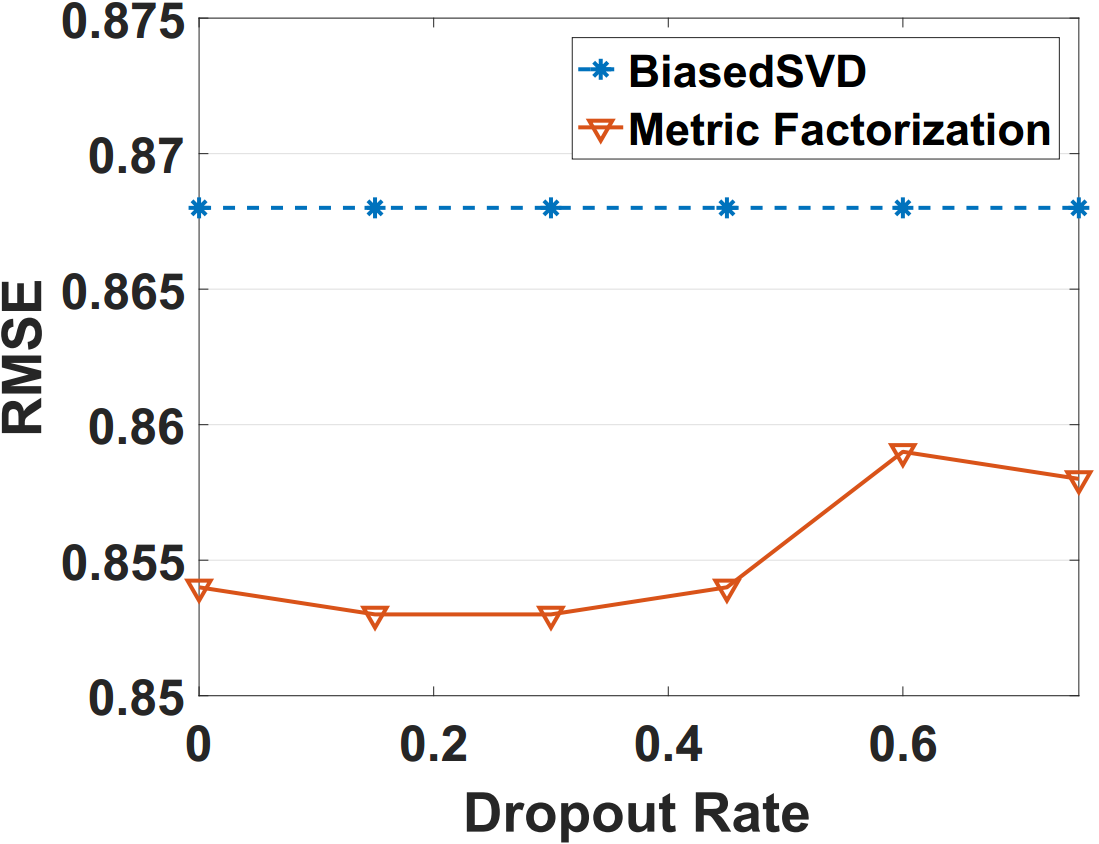}
\centering{(a). Rating Prediction}
\end{minipage}
\begin{minipage}[t]{4.1cm}
\includegraphics[width=4.1cm]{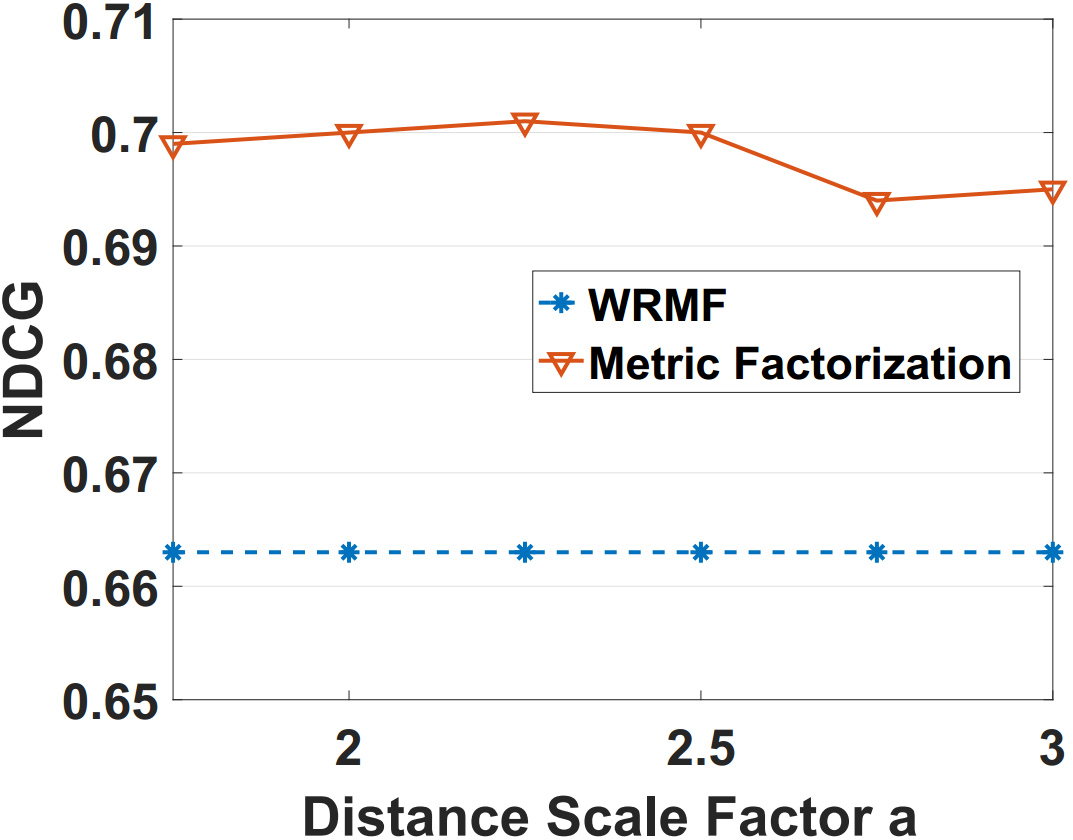}
\centering{(b). Item Ranking}
\end{minipage}
\caption{(a) RMSE with varying dropout rate; (b) NDCG with varying distance scale factor $a$. }
\label{fig:dropout}
\end{center}
\end{figure}

\subsubsection{Dropout Rate}
Figure \ref{fig:dropout}(a) shows the varying RMSE with the keep rate of dropout operation. It is observed that dropout does improve the performance of rating prediction. Nonetheless, the dropout rate is usually set to a small value as we do not want to drop useful information. Unfortunately, we find that the dropout operation does not favor the ranking based model.

\subsubsection{Distance Scale Factor $a$}
Since another distance scale factor $z$ is set to $0$, we only analyze parameter $a$. It controls the minimum distance that we would like to push the negative items away from the users. As shown the Figure 7(a), $a$ does not have as much influence as other parameters like $k$ and $l$. The performance on FilmTrust does not fluctuate much with $a$ being set to $1.75$ - $2.5$.

\section{Conclusion and Future Work}

In this paper, we propose the metric factorization which treats the recommendation problem from a novel perspective. Instead of modeling the user item similarity with dot product, we treat users and items as objects in the same multidimensional space and model their distances with Euclidean distance. We developed two variants of metric factorization to tackle the two classic recommendation tasks:rating prediction and item ranking. Experiments on a number of real world datasets from different domains showed that our model outperform conventional matrix factorization models and latest neural network and metric learning based approaches.

For future work, since metric factorization shares the similar structure as matrix factorization, we will explore its advanced extensions by accommodating temporal dynamics, side information such as item and user descriptions, review texts and even trust/social relationships. In this item ranking version of metric factorization, all negative entries of $R$ are included, we could explore some negative sampling strategy such as randomly sampling and dynamic negative sampling~\cite{Zhang:2013:OTC:2484028.2484126} to reduce the training samples






\bibliographystyle{IEEEtran}
\bibliography{reference}

\begin{thebibliography}{10}
\providecommand{\url}[1]{#1}
\csname url@samestyle\endcsname
\providecommand{\newblock}{\relax}
\providecommand{\bibinfo}[2]{#2}
\providecommand{\BIBentrySTDinterwordspacing}{\spaceskip=0pt\relax}
\providecommand{\BIBentryALTinterwordstretchfactor}{4}
\providecommand{\BIBentryALTinterwordspacing}{\spaceskip=\fontdimen2\font plus
\BIBentryALTinterwordstretchfactor\fontdimen3\font minus
  \fontdimen4\font\relax}
\providecommand{\BIBforeignlanguage}[2]{{%
\expandafter\ifx\csname l@#1\endcsname\relax
\typeout{** WARNING: IEEEtran.bst: No hyphenation pattern has been}%
\typeout{** loaded for the language `#1'. Using the pattern for}%
\typeout{** the default language instead.}%
\else
\language=\csname l@#1\endcsname
\fi
#2}}
\providecommand{\BIBdecl}{\relax}
\BIBdecl

\bibitem{Koren:2008:FMN:1401890.1401944}
Y.~Koren, ``Factorization meets the neighborhood: A multifaceted collaborative
  filtering model,'' in \emph{Proceedings of the 14th ACM SIGKDD International
  Conference on Knowledge Discovery and Data Mining}, ser. KDD '08.\hskip 1em
  plus 0.5em minus 0.4em\relax New York, NY, USA: ACM, 2008, pp. 426--434.

\bibitem{Rendle:2009:BBP:1795114.1795167}
S.~Rendle, C.~Freudenthaler, Z.~Gantner, and L.~Schmidt-Thieme, ``Bpr: Bayesian
  personalized ranking from implicit feedback,'' in \emph{Proceedings of the
  Twenty-Fifth Conference on Uncertainty in Artificial Intelligence}, ser. UAI
  '09.\hskip 1em plus 0.5em minus 0.4em\relax Arlington, Virginia, United
  States: AUAI Press, 2009, pp. 452--461.

\bibitem{Hu:2008:CFI:1510528.1511352}
Y.~Hu, Y.~Koren, and C.~Volinsky, ``Collaborative filtering for implicit
  feedback datasets,'' in \emph{ICDM}.\hskip 1em plus 0.5em minus 0.4em\relax
  Washington, DC, USA: IEEE Computer Society, 2008, pp. 263--272.

\bibitem{Salakhutdinov:2007:PMF:2981562.2981720}
R.~Salakhutdinov and A.~Mnih, ``Probabilistic matrix factorization,'' in
  \emph{Proceedings of the 20th International Conference on Neural Information
  Processing Systems}, ser. NIPS'07.\hskip 1em plus 0.5em minus 0.4em\relax
  USA: Curran Associates Inc., 2007, pp. 1257--1264.

\bibitem{he2017neural}
X.~He, L.~Liao, H.~Zhang, L.~Nie, X.~Hu, and T.-S. Chua, ``Neural collaborative
  filtering,'' in \emph{WWW}.\hskip 1em plus 0.5em minus 0.4em\relax
  International World Wide Web Conferences Steering Committee, 2017, pp.
  173--182.

\bibitem{Shi:2014:CFB:2620784.2556270}
Y.~Shi, M.~Larson, and A.~Hanjalic, ``Collaborative filtering beyond the
  user-item matrix: A survey of the state of the art and future challenges,''
  \emph{ACM Comput. Surv.}, vol.~47, no.~1, pp. 3:1--3:45, May 2014.

\bibitem{Ram:2012:MIS:2339530.2339677}
P.~Ram and A.~G. Gray, ``Maximum inner-product search using cone trees,'' in
  \emph{Proceedings of the 18th ACM SIGKDD International Conference on
  Knowledge Discovery and Data Mining}, ser. KDD '12.\hskip 1em plus 0.5em
  minus 0.4em\relax New York, NY, USA: ACM, 2012, pp. 931--939.

\bibitem{tversky1982similarity}
A.~Tversky and I.~Gati, ``Similarity, separability, and the triangle
  inequality.'' \emph{Psychological review}, vol.~89, no.~2, p. 123, 1982.

\bibitem{Koren:2009:MFT:1608565.1608614}
Y.~Koren, R.~Bell, and C.~Volinsky, ``Matrix factorization techniques for
  recommender systems,'' \emph{Computer}, vol.~42, no.~8, pp. 30--37, Aug.
  2009.

\bibitem{salakhutdinov2008bayesian}
R.~Salakhutdinov and A.~Mnih, ``Bayesian probabilistic matrix factorization
  using markov chain monte carlo,'' in \emph{Proceedings of the 25th
  international conference on Machine learning}.\hskip 1em plus 0.5em minus
  0.4em\relax ACM, 2008, pp. 880--887.

\bibitem{zhang2017deep}
S.~Zhang, L.~Yao, and A.~Sun, ``Deep learning based recommender system: A
  survey and new perspectives,'' \emph{arXiv preprint arXiv:1707.07435}, 2017.

\bibitem{DBLP:journals/corr/DziugaiteR15}
\BIBentryALTinterwordspacing
G.~K. Dziugaite and D.~M. Roy, ``Neural network matrix factorization,''
  \emph{CoRR}, vol. abs/1511.06443, 2015. [Online]. Available:
  \url{http://arxiv.org/abs/1511.06443}
\BIBentrySTDinterwordspacing

\bibitem{Wu:2016:CDA:2835776.2835837}
Y.~Wu, C.~DuBois, A.~X. Zheng, and M.~Ester, ``Collaborative denoising
  auto-encoders for top-n recommender systems,'' in \emph{Proceedings of the
  Ninth ACM International Conference on Web Search and Data Mining}, ser. WSDM
  '16.\hskip 1em plus 0.5em minus 0.4em\relax New York, NY, USA: ACM, 2016, pp.
  153--162.

\bibitem{Hsieh:2017:CML:3038912.3052639}
C.-K. Hsieh, L.~Yang, Y.~Cui, T.-Y. Lin, S.~Belongie, and D.~Estrin,
  ``Collaborative metric learning,'' in \emph{Proceedings of the 26th
  International Conference on World Wide Web}, ser. WWW '17.\hskip 1em plus
  0.5em minus 0.4em\relax Republic and Canton of Geneva, Switzerland:
  International World Wide Web Conferences Steering Committee, 2017, pp.
  193--201.

\bibitem{Weinberger:2009:DML:1577069.1577078}
K.~Q. Weinberger and L.~K. Saul, ``Distance metric learning for large margin
  nearest neighbor classification,'' \emph{J. Mach. Learn. Res.}, vol.~10, pp.
  207--244, Jun. 2009.

\bibitem{plant2014metric}
C.~Plant, ``Metric factorization for exploratory analysis of complex data,'' in
  \emph{Data Mining (ICDM), 2014 IEEE International Conference on}.\hskip 1em
  plus 0.5em minus 0.4em\relax IEEE, 2014, pp. 510--519.

\bibitem{Pan:2008:OCF:1510528.1511402}
R.~Pan, Y.~Zhou, B.~Cao, N.~N. Liu, R.~Lukose, M.~Scholz, and Q.~Yang,
  ``One-class collaborative filtering,'' in \emph{Proceedings of the 2008
  Eighth IEEE International Conference on Data Mining}, ser. ICDM '08.\hskip
  1em plus 0.5em minus 0.4em\relax Washington, DC, USA: IEEE Computer Society,
  2008, pp. 502--511.

\bibitem{He:2017:NCF:3038912.3052569}
X.~He, L.~Liao, H.~Zhang, L.~Nie, X.~Hu, and T.-S. Chua, ``Neural collaborative
  filtering,'' in \emph{Proceedings of the 26th International Conference on
  World Wide Web}, ser. WWW '17.\hskip 1em plus 0.5em minus 0.4em\relax
  Republic and Canton of Geneva, Switzerland: International World Wide Web
  Conferences Steering Committee, 2017, pp. 173--182.

\bibitem{dokmanic2015euclidean}
I.~Dokmanic, R.~Parhizkar, J.~Ranieri, and M.~Vetterli, ``Euclidean distance
  matrices: essential theory, algorithms, and applications,'' \emph{IEEE Signal
  Processing Magazine}, vol.~32, no.~6, pp. 12--30, 2015.

\bibitem{Jones:2011:CIR:2052138.2052389}
N.~Jones, A.~Brun, and A.~Boyer, ``Comparisons instead of ratings: Towards more
  stable preferences,'' in \emph{Proceedings of the 2011 IEEE/WIC/ACM
  International Conferences on Web Intelligence and Intelligent Agent
  Technology - Volume 01}, ser. WI-IAT '11.\hskip 1em plus 0.5em minus
  0.4em\relax Washington, DC, USA: IEEE Computer Society, 2011, pp. 451--456.

\bibitem{Amatriain:2009:ILI:1611644.1611670}
X.~Amatriain, J.~M. Pujol, and N.~Oliver, ``I like it... i like it not:
  Evaluating user ratings noise in recommender systems,'' in \emph{Proceedings
  of the 17th International Conference on User Modeling, Adaptation, and
  Personalization: Formerly UM and AH}, ser. UMAP '09.\hskip 1em plus 0.5em
  minus 0.4em\relax Berlin, Heidelberg: Springer-Verlag, 2009, pp. 247--258.

\bibitem{friedman2001elements}
J.~Friedman, T.~Hastie, and R.~Tibshirani, \emph{The elements of statistical
  learning}.\hskip 1em plus 0.5em minus 0.4em\relax Springer series in
  statistics New York, 2001, vol.~1.

\bibitem{srivastava2014dropout}
N.~Srivastava, G.~Hinton, A.~Krizhevsky, I.~Sutskever, and R.~Salakhutdinov,
  ``Dropout: A simple way to prevent neural networks from overfitting,''
  \emph{The Journal of Machine Learning Research}, vol.~15, no.~1, pp.
  1929--1958, 2014.

\bibitem{koren2008factorization}
Y.~Koren, ``Factorization meets the neighborhood: a multifaceted collaborative
  filtering model,'' in \emph{SIGKDD}.\hskip 1em plus 0.5em minus 0.4em\relax
  ACM, 2008, pp. 426--434.

\bibitem{Zhang:2017:AEH:3077136.3080689}
\BIBentryALTinterwordspacing
S.~Zhang, L.~Yao, and X.~Xu, ``Autosvd++: An efficient hybrid collaborative
  filtering model via contractive auto-encoders,'' in \emph{SIGIR}.\hskip 1em
  plus 0.5em minus 0.4em\relax New York, NY, USA: ACM, 2017, pp. 957--960.
  [Online]. Available: \url{http://doi.acm.org/10.1145/3077136.3080689}
\BIBentrySTDinterwordspacing

\bibitem{1423975}
G.~Adomavicius and A.~Tuzhilin, ``Toward the next generation of recommender
  systems: a survey of the state-of-the-art and possible extensions,''
  \emph{IEEE Transactions on Knowledge and Data Engineering}, vol.~17, no.~6,
  pp. 734--749, June 2005.

\bibitem{duchi2011adaptive}
J.~Duchi, E.~Hazan, and Y.~Singer, ``Adaptive subgradient methods for online
  learning and stochastic optimization,'' \emph{Journal of Machine Learning
  Research}, vol.~12, no. Jul, pp. 2121--2159, 2011.

\bibitem{Ricci:2010:RSH:1941884}
F.~Ricci, L.~Rokach, B.~Shapira, and P.~B. Kantor, \emph{Recommender Systems
  Handbook}, 1st~ed.\hskip 1em plus 0.5em minus 0.4em\relax New York, NY, USA:
  Springer-Verlag New York, Inc., 2010.

\bibitem{lemire2005slope}
D.~Lemire and A.~Maclachlan, ``Slope one predictors for online rating-based
  collaborative filtering,'' in \emph{Proceedings of the 2005 SIAM
  International Conference on Data Mining}.\hskip 1em plus 0.5em minus
  0.4em\relax SIAM, 2005, pp. 471--475.

\bibitem{Li:2017:NRR:3077136.3080822}
P.~Li, Z.~Wang, Z.~Ren, L.~Bing, and W.~Lam, ``Neural rating regression with
  abstractive tips generation for recommendation,'' in \emph{Proceedings of the
  40th International ACM SIGIR Conference on Research and Development in
  Information Retrieval}, ser. SIGIR '17.\hskip 1em plus 0.5em minus
  0.4em\relax New York, NY, USA: ACM, 2017, pp. 345--354.

\bibitem{dziugaite2015neural}
G.~K. Dziugaite and D.~M. Roy, ``Neural network matrix factorization,''
  \emph{arXiv preprint arXiv:1511.06443}, 2015.

\bibitem{guo2013novel}
G.~Guo, J.~Zhang, and N.~Yorke-Smith, ``A novel bayesian similarity measure for
  recommender systems,'' in \emph{Proceedings of the 23rd International Joint
  Conference on Artificial Intelligence (IJCAI)}, 2013, pp. 2619--2625.

\bibitem{shani2011evaluating}
G.~Shani and A.~Gunawardana, ``Evaluating recommendation systems,''
  \emph{Recommender systems handbook}, pp. 257--297, 2011.

\bibitem{deshpande2004item}
M.~Deshpande and G.~Karypis, ``Item-based top-n recommendation algorithms,''
  \emph{ACM Transactions on Information Systems (TOIS)}, vol.~22, no.~1, pp.
  143--177, 2004.

\bibitem{Zhang:2013:OTC:2484028.2484126}
W.~Zhang, T.~Chen, J.~Wang, and Y.~Yu, ``Optimizing top-n collaborative
  filtering via dynamic negative item sampling,'' in \emph{Proceedings of the
  36th International ACM SIGIR Conference on Research and Development in
  Information Retrieval}, ser. SIGIR '13.\hskip 1em plus 0.5em minus
  0.4em\relax New York, NY, USA: ACM, 2013, pp. 785--788.

\end{thebibliography}
%




\end{document}